\documentclass[useAMS,usenatbib,usegraphicx]{mn2e}

\title[Chandra X-ray Observations of the Young  Stellar Cluster NGC 6193]
{Chandra X-ray Observations of the Young Stellar Cluster \\
 NGC 6193 in the Ara OB1 Association}
\author[S.L. Skinner, S.A. Zhekov, F. Palla, and C.L.D.R. Barbosa]
{S.L. Skinner$^{1}$\thanks{E-mail:
skinners@casa.colorado.edu (SLS); szhekov@space.bas.bg (SAZ);
palla@arcetri.astro.it (FP); cassio@astro.iag.usp.br (CB)},
S.A. Zhekov$^{2,3}$, F. Palla$^{4}$, and C.L.D.R. Barbosa$^{5}$ \\
$^{1}$CASA,  Univ. of Colorado, Boulder, CO 80309-0389 USA \\
$^{2}$JILA,  Univ. of Colorado, Boulder, CO 80309-0440 USA \\
$^{3}$On leave from Space Research Institute, Moskovska str. 6, Sofia-1000, Bulgaria \\
$^{4}$INAF-Osservatorio  Astrofisico di Arcetri, Largo E. Fermi 5, 50125 Firenze, Italy  \\
$^{5}$Instituto de Astronomia, Geofisica e Ci\^{e}ncias Atmosf\'{e}ricas, Universidade
de S\~{a}o Paulo, Rua do Mat\~{a}o 1226, \\
05508-900 S\~{a}o Paulo, SP, Brazil} 
\begin{document}

\date{Accepted 2005 April 25. Received 2005 March 7; 
in original form 2005 March 7.}

\pagerange{\pageref{firstpage}--\pageref{lastpage}} \pubyear{2002}

\maketitle

\label{firstpage}

\begin{abstract}
A 90 ksec {\em Chandra} HETG observation of the young stellar cluster
NGC 6193 in the southern Ara OB1 association detected 43 X-ray         
sources in a 2$'$ $\times$ 2$'$ core region centered on the massive O stars
HD 150135 (O6.5V) and HD 150136 (O3 $+$ O6V). The cluster is dominated
by exceptionally bright X-ray emission from the two O stars, which are
separated by only 10$''$. The X-ray luminosity of HD 150136 is 
log L$_{\rm X}$ = 33.39 (ergs s$^{-1}$), making it one of the most
luminous O-star X-ray sources known. All of the fainter X-ray sources
in the core region have near-IR counterparts, but existing JHK photometry
provides little evidence for near-IR excesses. These core sources have
typical mean photon energies $\langle$E$\rangle$ $\approx$ 2 keV and 
about one-third are variable. It is likely that some are young low-mass
stars in the cluster, but cluster membership remains to be determined.
Grating spectra show
that the X-ray properties of HD 150135 and HD 150136 are similar, 
but not identical. Both have  moderately broadened unshifted emission
lines and their emission is dominated by cool plasma at kT $\approx$ 0.3 keV,
pointing to a wind-shock origin. However, the emission of HD 150136 is
slightly hotter and four times more luminous than its optical twin
HD 150135. We discuss the possibility that a radiative colliding wind
shock contributes to the prodigious X-ray output of the short-period
(2.66 d) spectroscopic binary HD 150136. A suprising result
is that the X-ray emission of HD 150136 is slowly variable on a timescale
of $<$1 day. The origin of the  variability is not yet known but the
observed behavior suggests that it is an occultation effect.
\end{abstract}


\section{Introduction}
The southern Ara OB1 association shows evidence of
recent star formation that may have been triggered
by  a supernova event (Herbst \& Havlen 1977, hereafter HH77;
Arnal et al. 1987). The remarkable young stellar cluster  
NGC 6193 lies near the center of Ara OB1 and was first
studied optically by Whiteoak (1963). A more
comprehensive optical study was undertaken by HH77
who determined 
the foreground reddening and derived a distance of
1.32 $\pm$ 0.12 kpc. They identified the higher mass
O, B, and A star cluster members, but lower mass members
have not yet been isolated. The cluster is undoubtedly
young with age estimates in the range $\sim$1.5 - 3.1 Myr
(Moffat \& Vogt 1973, HH77, V\'{a}zquez \& Feinstein 1992). 

The central part of the cluster  is dominated by two
luminous O-type stars HD 150135 and HD 150136, separated
by only $\approx$10$''$. The optical properties of these
two O stars are very similar, as summarized in 
Table 1. Their intense ionizing radiation may be 
influencing star-formation in the nearby RCW 108 
molecular cloud (Comer\'{o}n et al. 2005).

HD 150136 is unusual in several respects. It is 
a massive spectroscopic binary (SB2) 
consisting of two O stars in a close 2.662 day orbit
and has recently been classified as O3 $+$ O6V  (Niemela
\& Gamen 2005, hereafter NG05). The luminosity class
of the primary is not well-determined. 
For typical O star masses, the short
2.662 day period implies a  separation of just
a few stellar radii. At this close separation, wind-wind
interaction is expected. In addition, HD 150136 is a
strong centimeter radio source, and most of its radio
flux is  nonthermal (Benaglia et al. 2001). 
A fainter object (V = 9.0 mag) is visible 
$\approx$1.6$''$ north of HD 150136 (Mason et al. 1998), and 
we show here that this source is also visible
in the near-infrared (Sec. 3.3).
By comparison, HD 150135 is slightly fainter than 
HD 150136 in the optical and radio. Its binary status
is uncertain, but it has been classified
as a possible spectroscopic binary
(Garmany et al. 1980; Arnal et al. 
1988).

Although NGC 6193 has been studied  optically, there
are no previous pointed X-ray observations.  
However, a bright  X-ray source was
detected near the position of HD 150136 (= HR 6187) in the
{\em ROSAT} All Sky Survey (RASS). This X-ray source was 
identified with HD 150136 by Bergh\"{o}fer et al. (1996)
who obtained a luminosity in the 0.1 - 2.4 keV band
of log L$_{X}$ (ergs s$^{-1}$) = 33.05, making it the most
luminous O star detected in the RASS. This bright source was
also serendipitously detected in a  8460 s  {\em ROSAT} PSPC 
exposure (image rp900554n00) at a  $\approx$20$'$ off-axis
position with a count rate of  0.36 c s$^{-1}$, which is
40\% less than reported from the RASS detection.

The presence of a luminous (and possibly variable) X-ray
source in the {\em ROSAT} images at or near the HD 150136 
position motivated us to undertake a more detailed X-ray study 
of the NGC 6193 cluster core with the {\em Chandra} X-ray
Observatory (CXO). Our primary objectives were
to use  {\em Chandra's} high angular resolution to
determine the precise origin of the bright X-ray
emission detected by  {\em ROSAT} and to acquire a
grating spectrum capable of identifying the X-ray 
emission process. In addition, the sensitive
(90 ksec) {\em Chandra} observation provides the 
first high-quality X-ray image of the region
immediately surrounding the central O stars HD 150135
and HD 150136, allowing us to catalog the X-ray 
emitting population in the cluster core.

{\em Chandra} resolves the bright central X-ray source
into two components corresponding to separate but unequal
contributions from HD 150135 and HD 150136. Thus, the
luminous X-ray source detected by {\em ROSAT} is found
to be the superimposed contribution of the two
bright O stars lying  10$''$ apart.
Interestingly, the long {\em Chandra}
exposure shows that the emission of HD 150136 is slowly
variable on a timescale of  $<$1 day. 
Grating  X-ray spectra of HD 150135 and 150136
are similar (but not identical) and soft emission is
prevalent  in both stars, implying a wind shock origin.
We identify 43 X-ray sources in
the central $\approx$2$'$ $\times$ 2$'$ region 
of the cluster. All of these  have near-IR
counterparts and about one-third show  X-ray 
variability, thus being viable candidates for
low-mass cluster members.

\section[]{Observations}

\subsection{X-ray Observations}
The {\em Chandra} observation of NGC 6193 began on 27 June 2002
at 05:20 UT and ended on 28 June at 07:12 UT, yielding an exposure
live time of 90,337 seconds. The High Energy Transmission Grating
(HETG) was used along with the ACIS-S CCD detector in faint-timed
telemetry mode.  The nominal pointing position was
(J2000.0) RA = 16$h$ 41$m$ 18.89$s$, 
Decl. = $-$48$^{\circ}$ 45$'$ 39.3$''$, which is 16.7$''$
northwest of  HD 150136. Further information on
{\em Chandra} and its instruments can be found in 
Weisskopf et al. (2002).   

Data reduction used standard data processing (``pipeline'') 
products and   CIAO
\footnote{Further information on {\em Chandra}
Interactive Analysis of Observation (CIAO) software can
be found at http://asc.harvard.edu/ciao~.}
processing scripts supplied by the {\em Chandra} 
X-ray Center. Data were reduced using
CIAO vers. 3.0.2 and calibration data CALDB vers. 2.26.
Our post-pipeline
processing included steps to make use of observation-specific
bad pixel files, removal of faint point sources in the
regions used to extract grating spectra, destreaking
of the ACIS-S4 CCD, and energy filtering to reduce the
effects of low and high-energy background.

X-ray source detection was accomplished using the 
CIAO wavelet-based tool $wavdetect$ applied to 
full-resolution images
(0.49$''$ $\times$ 0.49$''$ pixels).
The images were first energy-filtered
to include only photons in the [0.5 - 7.0] keV energy
range, which reduces both soft and hard background 
emission. We ran $wavdetect$ using pixel scales
of 1,2,4,8, and 16 pixels, and compared results for
various values of the input parameter $sigthresh$, which
determines the false alarm probability. Images were
visually inspected for missed or spurious detections,
and we found that $sigthresh$ = 1.5 $\times$ 10$^{-5}$
produced good results. At this value, only one spurious
detection is expected in the 2.1$'$ $\times$ 2.1$'$ (256 $\times$
256 pixels) core region analyzed here (Fig. 1). 
The unabsorbed X-ray luminosity detection limit
corresponding to a 6 count on-axis detection in the 0th
order ACIS-S image 
is log L$_{\rm X}$ (0.5 - 7 keV) = 29.99 (ergs s$^{-1}$) at
the cluster distance of 1.3 kpc. This value was
determined from the Portable Interactive Mult-Mission
Simulator ({\em PIMMS})
\footnote{http://asc.harvard.edu/ciao/ahelp/pimms.html}, assuming an isothermal 
Raymond-Smith plasma with a characteristic temperature
kT = 2 keV and absoprtion column density N$_{\rm H}$ =
4 $\times$ 10$^{21}$ cm$^{-2}$. These spectral 
parameters are typical of X-ray sources in the
region (Sec. 3.2).

A total
of 43 X-ray sources were detected in the core region
(Table 2; Figure 1). 
Event lists were extracted for each source using the
elliptical source regions output by $wavdetect$. The
energy-filtered event lists 
were then used for further statistical analysis
of source properties (Sec. 3.2).

Background-subtracted first order grating spectra from the 
medium energy gratings (MEG) and high energy gratings (HEG)
were extracted for HD 150135 and HD 150136 using standard
CIAO threads. The roll angle of the observation was
favorable and provided adequate separation of the grating arms
of HD 150135 and HD 150136 on the ACIS-S detector. However,
we used a slightly narrower mask than the default value
to extract the grating spectra of each star in order to
avoid any possible cross-contamination. 
The CIAO tool {\em psextract} was used to extract
zeroth order spectra of selected bright sources (including
HD 150135 and HD 150136) along with background spectra from
adjacent source-free regions.

\subsection{Near-Infrared Observations}
We used the 2MASS all-sky data release
\footnote{See http://www.ipac.caltech.edu/2mass.}
to identify near-IR counterparts to X-ray sources
in the central cluster region (Table 2).
A few X-ray sources lacked 2MASS counterparts
but these were identified using more sensitive near-IR
images obtained with the 1.6 m Perkin-Elmer
telescope at the Observat\'orio do Pico dos Dias, Laborat\'orio 
Nacional de Astrof\'{\i}sica,  Brazil (Fig. 1). 
The observations were acquired 
on 23 July 2002 under photometric conditions ($<$0.7$''$ seeing)
with the CamIV near-IR camera. The camera uses  a
1024$\times$1024 pixels Hawaii HgCdTe detector with 0.24$''$
pixels, providing a 4$'$$\times$4$'$ field-of-view.
The central region of NGC 6193 was imaged in J, H, 
and a narrow  filter with a central wavelength 
$\lambda_{c}$=2.14 $\mu$m  ($\Delta\lambda$=0.018 $\mu$m).
Flux calibration was
achieved by observing near-infrared standard stars from 
Persson et al. (1998).  The data were processed in the
{\em IRAF} environment and the images were linearized, dark 
frame subtracted, flat-fielded, sky subtracted, and corrected
for bad pixels. Final images were constructed by co-adding
twelve exposures of 5 seconds each. Limiting magnitudes
of the final images were J = 19.5, H = 17.9, and
14.35 at 2.14 $\mu$m.

\section{Observational Results}

We present here the main  observational results 
including images, light curves and spectra.

\subsection{X-ray Image of the Central Cluster Region}

Figure 1 shows the {\em Chandra} ACIS-S 0th order
image of the central
2.1$'$ $\times$ 2.1$'$  region of NGC 6193 centered on
the optically bright O stars HD 150135 and HD 150136.
Although the 0th order image covers a larger region,
we restrict our analysis here to the central 
2.1$'$ $\times$ 2.1$'$ core where the {\em Chandra}
point-spread function is sharpest (thus providing
reliable source identification) and where the 
probability of cluster membership is high.

{\em Chandra} clearly detects both O stars. Furthermore, 
their emission is well-separated at {\em Chandra's}
high angular resolution and they are the two brightest
X-ray sources in the field.
$Wavdetect$  detected 42 X-ray
sources in this central region, and one additional
source (no. 26b) was found by visual inspection.
Table 2 summarizes the properties of these 43 sources
along with their IR/optical identifications. 
Near-IR counterparts were found for all 43 X-ray sources.
The two faintest sources recovered by $wavdetect$
have 4 counts each (sources 8 and 16), and are 
classified as possible detections. However,  both
have near-IR counterparts and the X-ray detections
are likely real.
A search of the {\em HST} Guide Star Catalog (GSC)
and USNO B1.0 catalogs revealed optical counterparts
within 1$''$ of the X-ray positions for 11 sources.
Four of the GSC counterparts are classified as
non-stellar. Included among the stellar identifications
are HD 150135, HD 150136, CD$-$48 11069 (B0-1), 
CD$-$48 11071 (B0V), and the V = 12.3 mag star NGC 6193-9.

\subsection{Global X-ray  Properties}

To obtain a quantitative measurement of variability,
we computed the K-S statistic from the unbinned 
energy-filtered event list of each source.
Further information on the K-S statistic can
be found in Press et al. (1992).
Fourteen sources were found to be variable (Table 2),
based on the criterion that they have a probability
of constant count rate P$_{const}$ $\leq$ 0.05.
Two of the four OB stars are variable. These are
HD 150136 (P$_{const}$ = 1.4 $\times$ 10$^{-8}$)
and CD$-$48 11069 (P$_{const}$ = 0.02). A 
moderate X-ray flare was detected in source no. 25,
which has a K = 11.4 mag 2MASS counterpart.
Since short-term ($\sim$1 day) variability is 
rarely detected in extragalactic AGNs, it
is likely that that most of the twelve variable sources
other than  HD 150136 and CD$-$48 11069 are
stellar.

To distinguish between soft and hard sources, the
mean photon energy $\langle$E$\rangle$ of each source 
(Table 2) was computed from energy-filtered event lists. 
Figure 2 is a histogram showing
the distribution of $\langle$E$\rangle$. Most sources
have $\langle$E$\rangle$ $\approx$ 1.8 - 2 keV, which
is typical of coronal emission from low-mass stars.
Three of the massive OB stars show a much
softer photon energy distribution that is suggestive of
a different X-ray emission process, probably related to 
shocks in their winds. These are HD 150135, HD 150136, 
and CD$-$48 11071. The latter star has spectral type
B0V and was the softest source detected in the central
region  with  $\langle$E$\rangle$ = 0.87 keV.

An interesting exception is the early B star CD$-$48 11069 
(source 26a), which shows a harder energy distribution 
$\langle$E$\rangle$ =  2.09 keV than the other three OB stars
as well as variable X-ray emission.
It has been variously classified as B0-1 V or B0 IV
(Reed \& Beatty 1995), and radial velocity variations (Arnal et al.
1988) indicate that it is a binary. It is thus possible
that the X-rays are due to a lower mass companion.

Except for the two bright O stars, only five other
X-ray sources in Table 2 have $>$100 counts 
(sources 5, 14, 25, 26a, 34). Thus, spectral
fitting is in most cases not practical because
of insufficient counts. However, we did fit the
CCD spectra of the above five sources with simple
absorbed one-temperature (1T) and two-temperature 
(2T) $vapec$ optically thin plasma models 
in XSPEC\footnote{The XSPEC spectral analysis
package is  developed and maintained by HEASARC at 
NASA's Goddard Space Flight Center. See
http://heasarc.gsfc.nasa.gov/docs/xanadu/xandu.html
for additional information.} version 11.3.1.
The fit parameters have large uncertainties but
acceptable 2T fits were obtained with a cool
component at kT$_{1}$ $\approx$ 0.1 - 0.4 keV and 
a hotter component at kT$_{2}$ $\approx$ 1.6 - 4.5 keV.
The highest temperatures were inferred for 
sources 14 and 25, both of which are variable and
thus likely to be magnetically-active low-mass
stars. The mean (median) absorption from 2T fits of
these five stars was N$_{\rm H}$ = 
5.0 (3.5) $\times$ 10$^{21}$ cm$^{-2}$, with a range of
a factor of $\sim$3. This absorption is similar to that
determined for the much brighter source HD 150136 (Table 3).   
For the B0-1 star CD$-$48 11069 noted above, a 2T
spectral fit gives log N$_{\rm H}$ = 21.2 (cm$^{-2}$) 
and a dominant hot component at kT $\approx$ 2.5 keV.

\subsection{Infrared Properties}
Both 2MASS and Pico dos Dias (PD) near-IR images 
show that the field near the cluster center 
($l$ = 336.7$^{\circ}$, $b$  = $-$1.57$^{\circ}$)
is crowded (Fig. 1). As a result, confusion flags are
set in one or more 2MASS bands for all but
10 of the sources listed in Table 2. We thus
provide here only a brief summary of the IR
properties and defer a complete discussion 
until  higher resolution images become available.

The near-IR colors of those
10 2MASS sources other than the two central O
stars without confusion flags at 
J, H, or K are  plotted in Fig. 3.
We also show the
colors of 4 2MASS sources with a confusion flag
set only at J band (not at H or K), but with
J magnitudes that were still in good agreement
($\Delta$J $\leq$ 0.2 mag) with Pico dos Dias 
photometry. The 14 sources in Fig. 3 range from
K = 7.91 (source 41 = CD $-$48 11071) to
K = 13.16 (source 2), with a mean (median)
K = 11.36 (11.49).   
Of the 14 sources in Fig. 3, only one
shows a  near-IR excess. This is source 30,
which is associated with a K = 11.19 2MASS 
counterpart and a R = 13.5 mag {\em HST}
Guide Star Catalog (GSC) source. The 
GSC classifies the object as non-stellar.

Thus, the most reliable JHK
data presently available  suggest that the
fraction of X-ray sources with near-IR excesses
is quite low. However, higher spatial 
resolution near-IR photometry is needed to 
obtain reliable colors for closely spaced
sources  along with mid-IR
photometry to search for cooler dust
emission that might be present.

The Pico dos Dias images   confirm
a previously detected
optical source (V = 9.0) located just 
north of HD 150136 (Mason et al. 1998).
This source is clearly visible
in our 2.14$\mu$ image (Fig. 4) at an offset of
$\approx$1.7$''$ north of HD 150136. We estimate
from Pico dos Dias photometry that it is 
$\approx$2.4 mag fainter than HD 150136 at 
2.14 $\mu$m, but higher angular resolution will be
needed to clearly separate the two. If this star is
at the same distance as HD 150136 and  has similar
A$_{V}$, then  it would correspond to an early B
spectral type for a main sequence object. No
significant X-ray emission is seen from this
star, which is in the wings of the {\em Chandra}
PSF of HD 150136.

\section{The Massive O Stars HD 150135 and HD 150136}

Although the two closely-spaced O stars HD 150135 and 
HD 150136 have similar optical properties, their X-ray
properties show differences, as discussed below.

\subsection{L$_{X}$/L$_{bol}$ }

The unabsorbed X-ray luminosity (L$_{X}$) of
HD 150136 is about 4 times larger than HD 150135
(Table 1). The high L$_{X}$ of  HD 150136 is
exceptional, making it one of the brightest (if
not {\em the} brightest) X-ray sources in the 
O star class.   But, the HD 150136 SB2 system
has  a larger bolometric 
luminosity  (L$_{bol}$) than HD 150135
by virtue of its earlier O3 primary and
O6V companion. As a result, the 
L$_{X}$/L$_{bol}$ ratios of HD 150135
and HD 150136 are nearly identical (Table 1).

The RASS study of OB stars by 
Bergh\"{o}fer et al. (1996, 1997) established a
correlation 
log L$_{X}$ = 1.13~log L$_{bol}$ $-$ 11.89,
with a standard deviation of 0.40 dex
for stars with log L$_{bol}$ (ergs s$^{-1}$)
$>$ 38. However,
many OB stars were not detected in the short
RASS exposures, so the mean  L$_{X}$ deduced
from RASS data is probably too high. A similar
correlation was found in {\em Einstein } data
for O stars by Sciortino et al. (1990). As
shown in Figure 5, the (L$_{X}$, L$_{bol}$) values 
for HD 150135 and HD 150136 are well within the
scatter of  the RASS sample, which is quite
large. Thus, the (L$_{X}$, L$_{bol}$) values for
these two  O stars do not seem
unusual, but  the high L$_{X}$ of 
HD 150136 is extreme for an O star.

\subsection{X-ray Light Curves of HD 150135 \\
            and HD 150136}

Figure 6 shows the 0th order ACIS-S light curves of
both HD 150135 and HD 150136. We have considered
the effects of pileup, which occurs when two or
more photons are detected as a single event.
Based on {\em PIMMS} count
rate simulations using fluxes determined from the 
MEG1 spectra (unpiled), we estimate that 0th order 
pileup was $\approx$9\% in HD 150136 and $\approx$1\%  
in HD 150135. The   low pileup level in HD 150135 is
negligible and the moderate level in HD 150136 will
cause the count rate to be slighly underestimated in
the 0th order light curve  but does not significantly
affect our results. The grating data are unaffected.

No significant variability
was detected in HD 150135, but HD 150136 is clearly variable.
Its count rate dropped slowly by about 38\% during the first 39 ksec
of the observation. It then
increased slowly back to levels near the initial value.
The first order MEG1 light curve shows similar behavior.
In addition, we extracted a light curve using
a small rectangular extraction region   of size 6 $\times$ 3
pixels (2.9$''$ EW $\times$ 1.5$''$ NS) centered on 
HD 150136. This region excluded the faint optical/IR source
located 1.7$''$ to the north, which in any case is not clearly 
detected in the X-ray images. The X-ray light curve
variability is still present, implying that the
faint optical/IR companion  is not the cause of
the X-ray variability.
A K-S analysis of the 0th order light curve of 
HD 150136 using events in the 0.5 - 5 keV range
gives a probability of constant count rate 
P(const) = 2.4 $\times$ 10$^{-8}$. 

To further investigate the nature of the variability,
we extracted energy-filtered light curves of HD 150136
in soft (0.3 - 1.5 keV) and hard (1.5 - 5 keV) bands.
We found no significant variation in the hardness ratio
during the light curve dip. In addition, we
extracted and fitted 0th order spectra of HD 150136
during the light curve dip (t = 24 - 48 ksec) and
after the dip (t = 55 - 90 ksec). Fits of these
spectra with two-temperature optically thin plasma
models show no significant  differences in temperature
or absorption. However, the total emission measure 
(sum of both temperature components)  during
the light curve dip is only 27\% of that in the
post-dip spectrum.

The above analysis shows that the X-ray variability
in HD 150136 was linked to a drop in the emission 
measure and was not accompanied by any significant
temperature change. This suggests that the variability
may have resulted from a partial obscuration of the 
X-ray emitting region, as discussed further in Section 5.8.

\subsection{X-ray Spectra of HD 150135 and HD 150136}

Figure 7 shows the 0th order spectra of
HD 150135 and HD 150136, and Figures 8 - 9
show their MEG1 spectra.
We have analyzed the 0th order and 1st order 
HETG spectra of both HD 150135 and HD 150136
using a variety of different models. These
include discrete temperature 
models with one (1T), two (2T), and three (3T) 
temperature components, and differential 
emission measure (DEM) models.  
Spectra were rebinned to a minimum of 20 counts
per bin for spectral fitting. All models
included an absorption component based on
Morrison \& McCammon (1983) cross sections.

Discrete temperature models were based on
the $bvapec$ module in XSPEC, which utilizes
the recent APED atomic data base and includes 
a  $velocity$ fit parameter that accounts for
line broadening. The DEM was constructed
using a modified version of the XSPEC
Chebyshev polynomial model  $c6pvmkl$ 
(Lemen et al. 1989), as discussed below.

Overall, our most robust results are based on
fits of the MEG1 spectrum of HD 150136, which
provides good spectral resolution and high
signal-to-noise (S/N) data. Measurements
of line properties of some lines were also
possible using HEG1 spectra of HD 150136 (Table 4),
which provides higher spectral resolution
than MEG1 at lower S/N. 
Because of its lower flux, our spectral
fits of  HD 150135 are restricted to
0th order and MEG1 spectra only.

\subsubsection{X-ray Temperatures}
Acceptable fits of 0th order spectra can 
be obtained using 2T models, as summarized
in Table 3. The spectra of
both  HD 150135 and HD 150136 are quite soft
with characteristic temperatures of kT $\approx$
0.2 - 0.3 keV. The MEG1 spectrum of
HD 150135 can be acceptably fit with a 1T
model having kT = 0.3 keV, and the 2T model
(Table 3) provides only a slight improvement.
In contrast, the 2T model provides a significant
improvement over 1T in HD 150136. Further modest
fit improvement can be obtained with a
3T model in HD 150136, which includes  a
third temperature component at kT $\approx$ 1.6 keV,
but  only 9\% of the total emission measure is 
attributed to this component.

\subsubsection{Emission Measure Distribution}

The DEM distribution of HD 150136 based on 
modeling of its MEG1 spectrum (8780 net counts) is shown
in Figure 10. The DEM was determined by a global
fit of the spectrum using  a modified 
version of the $c6pvmkl$ Chebyshev polynomial algorithm
in XSPEC that incorporates recent
APED atomic data. This global fitting strategy
takes the underlying continuum into account
in  reconstructing the DEM. Line broadening was
modeled  by applying the Gaussian
smoothing function $gsmooth$. Abundances were
varied and converged to values close to those
obtained for the 2T model (Table 3), with 
an inferred iron abundance Fe = 0.32 $\pm$ 0.09
($\pm$1$\sigma$) solar.

As seen in Figure 10, the DEM of HD 150136
is quite soft, with a single peak  near
kT $\approx$ 0.2 keV. The precise value of
the peak is somewhat uncertain, but is 
very  likely below 0.4 keV. We find no evidence
of a second hotter emission peak but some 
plasma up to at least kT $\approx$ 1 keV is 
inferred from the DEM profile.

We also attempted to reconstruct the DEM of 
HD 150135 using the Chebyshev polynomial method
applied to its MEG1 spectrum (1100 net counts).
The inferred DEM is very similar to that of 
HD 150136, showing a single soft peak below
kT $\approx$ 0.4 keV, with almost all fits
peaking at kT $\approx$ 0.2 - 0.3 keV.
One discernible difference is that the 
DEM of HD 150135 seems to be more sharply
peaked, with less fractional contribution 
from hotter plasma at temperatures above
the peak. This is consistent with the 
results of our discrete temperature models,
which also indicate a slightly cooler plasma
distribution in HD 150135.

\subsubsection{Emission Line Properties}

The emission line properties of  HD 150135
and HD 150136 are quite similar. Both stars
show broadened lines with typical broadening
of FWHM $\approx$2400 - 2500 km s$^{-1}$ (Table 3),
no significant line centroid shifts, and little
or no asymmetry. 
Table 4 summarizes the line properties of
HD 150136, which are more reliably determined
than for the fainter line emission of HD 150135.
Even so, the O VII line in HD 150136 is faint
and the uncertainties in its width and centroid
are larger than for the other lines. 
A notable result is that the higher temperature
S XV and Si XIV lines were
detected in HD 150136 (Fig. 9), but not in 
HD 150135 (Fig. 8).
MEG1 spectral simulations show that if the temperature
structure of HD 150135 were the same as that
of HD 150136, these two lines should have been
detected in HD 150135 in spite of its lower
count rate. We thus  conclude that the
plasma of  HD 150136 is slightly hotter than that
of HD 150135, as already suggested by discrete
temperature models.

\subsubsection{He-like Triplets}

Observed fluxes were measured for the resonance ($r$),
intercombination ($i$), and forbidden ($f$)
line components in He-like triplets of
HD 150136 using 1st order grating spectra. The flux 
ratios $R$ = $f/i$ and $G$ = ($f$ $+$ $i$)/$r$ are
given in Table 4, and the Si XIII triplet is
shown in Figure 11.  To measure the fluxes of
individual components, He-like triplets were
fitted with 3-component Gaussian models plus
a constant term to account for the underlying
continuum. To reduce the number of free parameters
in the fit,  the ratios of  the 
wavelengths of the $r$, $f$, and $i$
lines were held fixed during the 
fitting process at their laboratory values.  
In addition, line broadening was
taken into account. The line width was allowed
to vary during the fit, but the widths of all
three components were constrained to be equal.

The $G$ ratio is sensitive to temperature and
the $R$ ratio is sensitive to the electron
density ($n$) and stellar UV radiation field 
(Gabriel \& Jordan 1969). If $\phi$ is the
photoexcitation rate from the 1$s$2$s$~$^{3}S_{1}$ $f$
level to the 1$s$2$p$~$^{3}P_{1}$ $i$ level,
then $R$ = $R_{0}$/[1 $+$ ($\phi$/$\phi_{c}$) $+$ ($n$/$n_{c}$)],
where $\phi_{c}$ and $n_{c}$ are the critical photoexcitation
rate and critical density (Blumenthal, Drake, \& Tucker 1972;
Porquet et al. 2001).
If photoexcitation and collisions are both negligible,
then $R$ $\rightarrow$  R$_{0}$, where  R$_{0}$ is
referred to as the low-density limit.
Either high electron densities
or strong photoexcitation can lead to  $R$ $<$  R$_{0}$.

We obtain $R$ $<$  R$_{0}$ for all He-like
triplets measured in HD 150136, with S XV
being a possible exception (Table 4). As 
discussed above, the suppressed $R$ values could
be due to high electron densities or intense UV 
radiation (or both). However, it is unlikely
that the suppression can be explained by high
densities. As an example, consider Ne IX $\lambda$ 13.447 \AA,
for which $n_{c}$ $\sim$ 10$^{11}$ cm$^{-3}$
(Porquet \& Dubau 2000). Assuming a  spherically
symmetric homogeneous solar-abundance wind,
mass loss parameters as in Table 5, and a 
conventional wind velocity profile
$v(r)$ = $v_{\infty}$[1 $-$ (R$_{*}$/r)]$^{\beta}$
with $\beta$ = 1, densities $n$ $\ga$ $n_{c}$ are anticipated 
only at radii  $r$ $\la$ 1.2 R$_{*}$ for an O3V 
primary, and the corresponding radius for O3I is 
about 30\%  larger.
However, it is clear that the Ne IX line cannot
originate at $r$ $\la$ 1.2 R$_{*}$ because the 
radius of optical depth unity for O3V at the 
Ne IX  line energy 0.92 keV is
R$_{\tau = 1}$(0.92 keV) $\approx$ 5.5 R$_{*}$,
and about five time larger for O3I. To
estimate  R$_{\tau = 1}$, we have used
solar abundance cross-sections (Balucinska-Church
\& McCammon 1992; Anders \& Grevesse 1989) along with the
O3 stellar parameters in Table 5. Thus, the 
optical depth constraint places the  region of
line formation for Ne IX outside the radius at
which density suppression would be important.
We emphasize that this conclusion rests on the
assumption of a  spherical homogeneous wind.
If the wind is clumped (Sec. 5.4), then 
R$_{\tau = 1}$ is reduced and line formation 
could occur at smaller radii  than inferred
above.

It is more likely that the low R values
result from photoexcitation by the strong
UV field of the  O3 $+$ O6V system. The distance
at which UV effects become significant is
measured by the critical radius R$_{c}$,
where the photoexcitation rate from the upper
level of the $f$ transition equals the
spontaneous decay rate from that same level.
Triplets forming at $r$ $<$  R$_{c}$ will
have UV-suppressed R ratios.
Values of R$_{c}$ for O3V are given in Table 4
for each He-like triplet.

Figure 12 shows R$_{\tau = 1}$ and R$_{c}$
values for  He-like triplets in HD 150136. 
In general, UV-suppressed triplets could
form at radii R$_{\tau = 1}$ $<$ $r$ $<$  R$_{c}$,
as shown graphically by the solid lines in
Figure 12. But, in HD 150136 the R ratios are
strongly suppressed and it is more likely
that $r$ $<<$  R$_{c}$. In any case,
it is apparent from Figure 12 that triplets with 
higher maximum line power temperatures
such as S XV and Si XIII are formed closer
to the star than lower temperature lines
such as Ne IX and O VII.

\subsubsection{Abundances}

The MEG1 spectrum of HD 150136 is of sufficient quality
to derive basic abundance information (Table 4), The strongest
result is that  Fe is underabundant with respect
to the solar photospheric value by a factor of
$\approx$4 - 5, and quite likely
Ne as well. The Fe underabundance is also clearly
seen in the MEG1 spectrum of HD 150135. Low
Fe abundances have also been reported for other 
massive young stars such as Trapezium OB stars
(Schulz et al. 2003). For HD 150136, O and S
abundances have the largest uncertainties.

\section{Discussion}

We summarize here the stellar properties of HD 150135
and HD 150136 that are relevant to the X-ray analysis,
and discuss the implications of the new {\em Chandra}
results for our understanding
of the X-ray emission in massive young stars.

\subsection{The Short-Period Binary System HD 150136}
The orbital period of the O3 $+$ O6V
system is 2.662 $\pm$ 0.002 days, the orbit is nearly 
circular ($e$ = 0.03 $\pm$ 0.02), and the
inclination is probably low since optical 
eclipses have not been seen (NG05).
The luminosity class of the O3 primary is not
well-determined, and we consider both O3I and O3V.
Adopting the O star parameters listed in Table 5,
the orbital semi-major axis from Kepler's 3rd 
law is $a$ = 0.165 AU for
a O3V primary and  $a$ = 0.173  AU for
O3I.  In units of O3 star radii,
this separation equates to only 
2.1 - 2.7 R$_{*}$. Thus, this remarkable
O $+$ O system is nearly in contact.
As a result of the  close separation, interaction
of the O3 and O6V winds seems unavoidable and
may lead to X-ray emission from a colliding wind
(CW) shock, as discussed further below.

\subsection{Is HD 150135 a Binary?}
The binary status of HD 150136 is firmly
established (NG05) but 
the evidence for binarity in HD 150135 
is not as conclusive.
Garmany et al. (1980) classified it as a possible
SB1 based on large radial velocity differences
in three spectra. Arnal et al. (1988)
obtained 17 radial velocity measurements over
10 days and classified it as a possible SB2.

The radio continuum emission of HD 150135
(and HD 150136) is
unusually strong for a star of its spectral
type, giving further reason to suspect binarity.
The flux density of HD 150135 increases with frequency 
with a 3.5 cm value   S$_{3.5 cm}$ = 
0.28 $\pm$ 0.03 mJy (Benaglia et al. 2001).
If this flux is attributed entirely to
free-free wind emission, then the inferred
mass loss rate at d = 1.3 kpc is
log $\dot{M}$ = $-$5.4 M$_{\odot}$ yr$^{-1}$.
This value is about an order of magnitude
larger than determined  for other O6.5V stars
(Garmany et al. 1981; Lamers \& Leitherer 1993).
It is thus possible that some of the radio
flux is nonthermal, as is  the
case for HD 150136.

The above results point toward binarity in
HD 150135, but nothing is known about the
putative companion. The {\em Chandra}
data do not  show any telltale hints  of
binarity such as variability or hard X-ray
emission that might arise in a colliding wind shock.
Its L$_{X}$/L$_{bol}$ ratio is high, but
within the scatter of the RASS sample and
nearly on the regression line for O stars
determined by  Sciortino et al. (1990) (Fig. 5).

\subsection{Wind Parameters}
An estimate of the terminal
wind speed is available for HD 150136
(Prinja et al. 1990) but not for HD 150135.
Both stars are radio continuum sources
(Benaglia et al. 2001), but HD 150136
contains a nonthermal contribution, as may
HD 150135 (Sec. 5.2).
Thus, estimates of the mass loss
rates based on radio continuum fluxes are
likely to be incorrect. In the absence of
reliable observational data, we have 
adopted the mass loss parameters given
in Table 5, which are representative of
O3 and O6V stars in general.

\subsection{Line Properties and Clumped Winds}
The line widths of HD 150136  in Table 4 reveal
interesting trends 
 The widths of all lines
are similar, with a mean $\overline{ \rm FWHM}$
= 2100 km s$^{-1}$. Then, $\overline{\rm  HWHM}$ = 
1050 km s$^{-1}$, or  $\overline{\rm  HWHM}$
$\approx$  v$_{\infty}$/3 (Table 5). Thus, the lines are
broadened, but not nearly as broad as
might be expected if they were formed
in the outer wind where the wind 
is at terminal speed. A similar
situation exists for HD 150135, whose
mean line width as determined from MEG1
spectra is nearly identical to that found
in HD 150136. Recent {\em Chandra} grating
observations of other young O stars such
as $\sigma$ Ori AB (O9.5V) also show very moderate
line broadening with HWHM $\approx$ $v_{\infty}$/4
(Skinner et al. 2004).

If one attributes the  broadening to X-ray line
formation in an outflowing stellar wind
(Owocki \& Cohen 2001), then a 
$\beta$ = 1 wind acceleration profile
(Sec. 4.3.4) gives 
$v(r)$ = $\overline{\rm  HWHM}$ = 1050 km s$^{-1}$
at  $r$ $\approx$  1.5 R$_{*}$. Comparing
against Figure 12, it is apparent that
higher temperature 
lines could form at such small radii but
cooler lines such as Ne IX and O VII  are
expected to form  further out because of
their larger R$_{\tau = 1}$ values.
Thus, there is an apparent contradiction.
Moderately broadened lines suggest that
the lines are formed in the wind acceration
zone close to the star, but optical depth
calculations require  larger
formation radii for cooler lines, where
the wind should be  at or near its terminal speed.

The above inconsistency may be due to our
imprecise knowledge of the mass loss parameters,
wind opacities, and wind structure. The
wind acceleration profile is not empirically
determined, and the mass-loss rate, which
enters into the  optical depth unity calculation,
is poorly-known. Lastly,
the assumption of a spherical homogeneous
wind may be overly simplistic. It has
been suggested that O star winds
may be clumped and that clumping could
account for discrepancies in O star
mass loss rates determined by different
methods  (Fullerton et al. 2004; 
Repolust et al. 2004).
If the wind is clumped then the optical
depth at a given line energy would be
reduced and line photons could escape
from  smaller radii than indicated
by Figure 12, resulting in less broadening
than expected for a homogeneous wind
(Owocki \& Cohen 2001).

\subsection{Colliding Winds in HD 150136}

The close separation in the HD 150136 O3 $+$ O6
system virtually assures wind-wind interaction. 
We would thus like to know if some or all of
its prodigious X-ray emission  could be due a colliding
wind shock, which is expected to form in the
harsh  radiation environment between
the two O stars. There are indeed signs that 
the X-ray emission of HD 150136 may be influenced by
binarity. As we have seen, its X-ray emission
is slightly hotter and a factor of four more 
luminous than its optical twin HD 150135.
In addition, its strong nonthermal radio
emission may signal shock processes.
Because of the close binary separation, 
the radiation-hydrodynamics problem
is coupled and is  challenging to model
numerically.

\subsubsection{Non-adiabatic Colliding Wind Shocks}

The temperature of shocked gas at the shock 
interface in an  {\em adiabatic} colliding wind 
shock is proportional to the square of the 
pre-shock wind velocity component normal to the
interface (eq. [1]. of Luo et al 1990). For
wind speeds typical of O stars  (Table 5), X-ray
temperatures of several keV are expected 
and have been observed in such wide systems
as the 7.9 year WR $+$ O binary WR 140 (Zhekov \&
Skinner  2000).

In contrast, we  detect very little hot plasma 
above $\sim$1 keV in HD 150136 (Fig. 10). Because of
its closer separation, the winds will not
have reached terminal speeds before colliding
and radiative cooling will be important. As a 
result, colliding wind shock emission could be
present at lower temperatures than predicted in the
adiabatic case.

The  parameter used to determine whether a colliding
wind shock is adiabatic or non-adiabatic is the 
ratio $\chi$ = t$_{cool}$/t$_{esc}$, where
t$_{cool}$ is the characteristic cooling time
for shock-heated gas and t$_{esc}$ is the 
escape time for gas to flow
out of the intershock region. If  $\chi$
$\ga$ 1 then the system is adiabatic while
$\chi$ $<<$ 1 implies the system is non-adiabatic
and radiative cooling is important. Adopting
O3V wind parameters for HD 150136 (Table 5)
and a $\beta$ = 1 velocity law, 
we obtain $\chi$(O3V) = 0.2 (eq. [8] of 
Stevens et al. 1992, hereafter SBP92).
Thus, HD 150136 is a non-adiabatic system.

\subsubsection{Colliding Wind Simulations}

The numerical hydrodynamic simulation of colliding
winds in closely-spaced  non-adiabatic systems is 
complex. Cooling processes and wind acceleration 
must be taken into account. Also, radiative 
braking may be important, whereby the radiation
of one O star slows the wind of its companion
(Owocki \& Gayley 1995). The presence of a 
mathematical singularity at the stagnation 
point on the line-of-centers has hindered the
development of rigorous numerical hydrodynamic
models, as discussed by Myasnikov et al. (1998). 

Because of the above complexities, we have only
developed simplified hydrodynamic CW simulations
of HD 150136 to date.  We assume O3V $+$ O6V wind 
parameters, a $\beta$ = 1 velocity law, and 
the orbit solution of NG05. The simulations invoke
wind momentum balance and the  shock surfaces are 
assumed to follow the contact discontinuity (Luo
et al. 1990; SBP92), as expected for efficient
cooling. Cooling from an optically thin plasma 
was included using the approach of Myasnikov
et al. (1998), as well as cooling from
Comptonization (Myasnikov \& Zhekov 1993).

The simulations indicate that a CW shock model
alone cannot account for all of the observed
X-ray properties of HD 150136. Specifically,
the CW model underestimates N$_{\rm H}$ by
about an order of magnitude, and also
places a larger fraction of the emission
measure in hot plasma (kT $\geq$ 1 keV) than
expected from the observationally determined
DEM (Fig. 10).

The above results indicate that if CW shock 
emission is present, then it is supplemented 
by an additional cool component, which would most
likely originate in the O stars themselves.
X-ray emission from cool plasma in the shocked
line-driven winds of the individual O stars is
indeed expected (Lucy \& White 1980; Lucy 1982). 
To test this idea, we attempted to fit the MEG1
spectrum of 
HD 150136 using a hybrid 1T $+$ CW model, where
the isothermal 1T thin plasma component accounts
for cool shocked plasma in the O-star winds.

This hybrid model provides
a much better fit to the spectrum than does the
pure CW model, resulting in a 37\% decrease in
reduced $\chi^2$ ($\chi^2$/dof = 213/377).
The hybrid model satisfactorily
reproduces the overall spectral shape, observed flux, 
and N$_{\rm H}$, and converges to abundances that
are very close to those obtained in discrete 
2T fits (Table 3). The 1T component in the
hybrid model converges
to kT $\approx$ 0.3 keV, consistent with expectations
from cool shocked plasma in the O star winds.
Roughly half of the observed
flux could come from the O stars themselves.
However, we  note that our {\em a priori} estimate
of the emission measure from the CW shock is a 
few times larger than  needed to accurately
reproduce the MEG1 spectrum.  Several factors 
could account for this including inaccuracies
in the orbit solution or the assumed mass loss rates,
or oversimplifications in the CW hydrodynamic model.

The hybrid 1T $+$ CW model  offers a potential
explanation of the X-ray properties of HD 150136. 
But, additional refinements in the CW model
as well as a more accurate orbit solution and
determination of the luminosity class of the primary
will probably be needed to determine if the hybrid
model can be fully reconciled with the data.
Further refinements in the CW shock model will include
radiative braking and accounting for the effects of
orbital motion on the emission line properties.

\subsection{Magnetically Confined Winds}
X-ray grating spectra recently obtained
for several massive young stars have sparked
renewed interest in the possibility that
their winds might in some cases be magnetically confined.
This has been the result of the detection
of narrow emission lines and hot plasma
(kT $\ga$ 2 keV) in young objects such
as the O star $\theta^1$ Ori C (Schulz et al. 2000)
and the unusual B star  $\tau$ Sco (Cohen et al. 2003).
Narrow unshifted lines and hot plasma are 
usually associated with magnetically trapped
plasma and are difficult to explain in the context 
of classical (non-magnetic) wind shock theory.

The influence of a dipole magnetic field on
the outflowing wind of a luminous hot star
has been discussed by ud-Doula \& Owocki (2002).
The degree to which the wind is confined by
the field is measured by the confinement
parameter  $\eta$ =
(0.4~B$_{100}^{2}$R$_{12}^{2}$)/($\dot{\rm M}_{-6}$$v_{8}$)
where B$_{100}$ is the equatorial magnetic field 
surface strength in units of
100 G, R$_{12}$ is the stellar radius in units of 
10$^{12}$ cm, $\dot{\rm M}_{-6}$ is the mass loss
rate in units of 10$^{-6}$ M$_{\odot}$ yr$^{-1}$,
and $v_{8}$ is the wind speed in units of
10$^{8}$ cm s$^{-1}$. If $\eta$ $>$ 1 (strong confinement)
then the
wind is confined by the B field and is channeled
toward the magnetic equator, where the wind
components from each hemisphere collide to form
a magnetically confined wind shock (MCWS),  
accompanied by hard X-ray emission
(Babel \& Montmerle 1997). 
If $\eta$ $<$ 1 (weak confinement), the field
is opened by the dominant wind and no
hard X-rays are predicted.

Since HD 150135 and HD 150136 are massive young
stars, we ask if the above model could be 
relevant to their X-ray emission. It is clear
that we do not detect the hot (kT $\ga$ 2 keV)
plasma or the very narrow lines that the MCWS
model attempts to explain in other stars. Thus,
there is no compelling reason to invoke this model 
in the present case, at least in the strong
confinement limit where  $\eta$ $>$ 1. 
If this model is relevant at all for the 
O stars studied here, 
it must be in the weak confinement regime where
$\eta$ $<$ 1. For critical confinement ($\eta$ = 1)
the mass loss parameters in Table 5 (O3V) give 
B$_{crit}$ $\approx$ 400 G. Thus, surface
fields no larger than a few hundred Gauss 
would be expected for HD 150136. Similarly,
the magnetically confined wind model of 
Usov \& Melrose (1992) gives
a surface field estimate of $\approx$ 140 G
in the weak-field approximation.

\subsection{Magnetic Fields and Early  Evolution}

Previous studies of NGC 6193  gave age estimates
of $\sim$1.5 - 3.1 Myr   
(Moffat \& Vogt 1973; HH77; V\'{a}zquez \& Feinstein 1992).
We can now compare these  with estimates of the ages
of the central O stars based on updated evolutionary models
For this purpose, we use the grids of Schaller et 
al. (1992). Because of the complications of determining 
the photometric and spectroscopic properties of
the two O stars in the HD 150136 SB2 system, we
consider HD 150135 as a more reliable age indicator
(but keeping in mind that it may also be a binary). 
Its absolute magnitude is M$_{v}$ = $-$5.33 (HH77).
The O star parameters of Vacca et al. (1996) for
luminosity class V then give log T$_{eff}$ =
4.66,  log (L/L$_{\odot}$) = 5.727, and an
evolutionary mass M$_{\rm evol}$ = 56.6 M$_{\odot}$.
Using these stellar parameters, the Schaller et al. tracks
for metallicity Z = 0.020 and 60 M$_{\odot}$ 
give an age of $\sim$1 Myr. A similar age is expected
for HD 150136 because of its close association with
HD 150135.

The age of $\sim$1 Myr deduced for HD 150135
and, by association, for HD 150136 is a few 
times larger than the estimated age of 
$\sim$0.3 Myr for Trapezium OB stars,
which may still be on the zero-age main
sequence (Schulz et al. 2003). It is
notable that several high-mass stars in the
Trapezium such as $\theta^1$ Ori C show signs
of magnetic behavior such as hot plasma,
very narrow  emission
lines, and periodic  X-ray variability
(Schulz et al. 2003, Feigelson et al. 2002,
Gagn\'{e} et al. 1997). In contrast, we
do not detect obvious X-ray magnetic signatures
in HD 150135 or HD 150136, although periodic
X-ray variability in the latter is not yet
ruled out due to limited time monitoring.
In the absence of any recognizable magnetic
X-ray behavior, it may  well be that any 
primordial magnetic fields that were initially
present in HD 150135 or HD 150136 have already
decayed to low levels. If that is the case,
then the estimated age of $\sim$1 Myr suggests
that the timescale for magnetic decay is 
quite short, and the possibilities for 
studying magnetic behavior in O stars may
be limited to the very youngest 
objects that are still on the zero-age 
main sequence.

\subsection{X-ray Variability in O Stars}

As discussed in Section 4.2, HD 150136 shows
slow X-ray variability in the form of a dip
in its light curve spanning at least $\sim$30 
ksec (Fig. 6). The variability was detected at
high statistical significance in both the 0th
order and MEG 1st order light curves. Simultaneous
light curves of HD 150135 lying 10$''$ away on the
same detector show no significant variability.
These factors leave little doubt that the X-ray
variability of HD 150136 is real. 

X-ray variability in O stars has been known since
early observations with the {\em Einstein} observatory, 
but its origin is not yet understood
and more than one mechanism may be involved.
Collura et al. (1989) detected significant variability
in three of twelve OB stars observed 
with  the {\em Einstein}
IPC. The variability occurred on timescales of a few
days with typical variability amplitudes of $\approx$17\%
- 34\%. We find a similar variability amplitude for
HD 150136, whose count rate fluctuated  $+$16\%/$-$22\%
about the mean. Lower amplitude variability at the
6\% level was reported from {\em ROSAT} observations
of the O4 supergiant $\zeta$ Puppis (Bergh\"{o}fer
et al. 1996b). Variability
was also detected in three O stars in the Orion Nebula
Cluster (ONC) observed by {\em Chandra} (Feigelson
et al. 2002).  These were $\theta^1$ Ori C (O6pe),
$\theta^1$ Ori A (O7), and $\theta^2$ Ori A (O9.5Vpe).
The latter star underwent a rapid flare. 

There are several possibilities for explaining X-ray
variability in O stars. These can be broadly
categorized as  (i) magnetic processes, (ii) changes
in the wind (e.g. density variations or time-variable
shock structures), and (iii) dynamical effects.
The latter category is quite broad and is meant
to include phenomena such as (partial) eclipses
of the X-ray emitting region by a companion,
changes in X-ray absorption caused by the 
wind or atmosphere of a companion moving into 
the line of sight, changes in orbital separation
in CW systems, and rotational modulation of
X-ray emitting structures on or near the star.

Intrinsic variability associated with magnetic  processes is
usually accompanied by high-temperature plasma
(kT $>>$ 1 keV) and/or rapid magnetic reconnection
flares. Dramatic temperature changes can occur on
timescales of minutes to hours in magnetic flares.
We do not detect high-temperature plasma in HD 150136,
nor do we see any significant temperature changes or
rapid impulsive flares. Thus, the existing data
for HD 150136 do not obviously point toward a 
magnetic origin for the slow variability. But
the temporal coverage is so far quite limited
and spans only 39\% of the 2.662 day orbit.
Observations spanning a longer time period
are needed.

Changes in the wind density at the base of the
wind were cited as a possible explanation of
the X-ray variability detected by {\em ROSAT}
in the O4 supergiant $\zeta$ Pup (Bergh\"{o}fer
et al. 1996b).  However, the
mechanism that would drive such density perturbations
is not known. Along  similar lines,
Feldmeier et al. (1997) have attributed the 
X-ray variability in O stars to the collision
of dense shells of gas in their winds.
This model predicts 
short X-ray variability timescales of $\sim$500 s.
This is much shorter than the variability timescale
of several hours observed in HD 150136, 
but one might argue that
variability timescales of $\sim$hours could
result from successive collisions of dense
shells (Fig. 19 of Feldmeier et al. 1997).
Apart from the timescale question, the high
X-ray luminosity of HD 150136 may be even 
more problematic for the shell-collision
picture. Typical X-ray luminosities predicted
by shell-collision simulations are log L$_{\rm X}$
$\approx$ 32.3 (ergs s$^{-1}$), with model-dependent
standard deviations of $\pm$0.4 to $\pm$0.6.  
This predicted L$_{\rm X}$ is
about an order of magnitude less than determined for
HD 150136, even after halving the total   L$_{\rm X}$
(Table 1) to account for two stars in the  system.
It thus remains to be shown that the shell-collision
mechanism can account for the very high X-ray luminosity
and slow variability of HD 150136. 

Finally, we consider dynamic effects. These are
indeed relevant because HD 150136 is known to
be a 2.662 d binary.
During our observation, the orbital phase of HD 150136 
ranged from $\phi$ = 0.1 - 0.5, where the primary is
in front at $\phi$ = 0.5 (NG05).
The dip in the light curve is centered near 
$\phi$ $\approx$ 0.27 (Fig. 6). At this phase, the system is
near quadrature, with the O3 primary approaching
the observer and the O6 secondary receding. 

In colliding wind systems, X-ray variability
can be induced by orbital motion. The intrinsic
luminosity of a CW shock is predicted to scale
inversely as the distance between the two
components (SBP92), but the orbit of HD 150136
is nearly circular ($e$ = 0.03; NG05) so no significant
variability due to changes in orbital separation is
expected. Also, a reduction in the 
observed X-ray flux can occur in colliding wind
binaries from increased wind absorption as 
the primary moves in front of the secondary.
Again, this does not explain the dip in the
HD 150136 light curve, which occurs near 
quadrature and not when the primary is in front.
To explain the light curve dip at quadrature
in terms of colliding wind geometry, one would
have to postulate that some of the CW shock 
emission on the downstream side of the secondary
is being shadowed by the secondary itself.

A possible clue to the origin of the X-ray
variability comes from the optical spectra of
NG05. They note that the radial velocity
changes in the He I absorption lines of HD 150136
do not follow the orbital motion of either the
primary or secondary.  Furthermore, they
detect an {\em additional} faint absorption
component redward of the  He I absorption lines
when the secondary of the SB2 system
has maximum positive radial velocity,
which occurs at phase $\phi$ = 0.25.
This may signal the  presence of a third
massive star in the system. We emphasize
that this putative third component in 
the spectroscopic system is {\em not}
the faint star lying 1.7$''$ to the north
(Fig. 4), whose position is well-known from
optical studies (Mason et al. 1998).

It may not be a coincidence that the 
appearance of the additional optical absorption
component occurs at $\phi$ = 0.25, where
the dip in the X-ray spectrum is also
observed (Fig. 6). If the additional
absorption feature is due to a massive third
component moving in front at 
$\phi$ = 0.25, then this third star (or its
wind) could be partially occulting the 
X-ray emitting region. Such an occultation
would be consistent with the observed 
decline in emission measure, without an
accompanying temperature change. However,
a physical connection between the X-ray
variability and the putative third star
remains quite speculative. It might
be argued that a physical connection is
unlikely, since it seems to require that
the orbit of the third star be synchronous
with that of the SB2 system. Clearly,
further observational work is needed to
determine if there is a connection. This
would include confirmation of the existence
of the putative third star, determination
of its spectral type and orbit, and 
extended time monitoring to search for
periodicity in the X-ray light curve.

A final possibility worth mentioning is that
the X-ray variability of HD 150136 might be
due to rotational modulation of an X-ray
emitting structure that is close to the
star. In that case, the observed variability
could result from self-occulation, rather
than from occultation by a putative third
star in the system. Rotational self-occultation
has been invoked to explain the X-ray variability
of $\theta^1$ Ori C (Gagn\'{e} et al. 2005).
Its variability is caused
by a change in emission measure without an
accompanying temperature change, which is
qualitatively similar to the behavior that we
see in HD 150136. The key question that
remains is whether the X-ray variability
in HD 150136 is periodic and if so, what is 
the period?

\section{Summary}

We have presented results of the first X-ray
observation targeted at the center of the young
stellar cluster NGC 6193 in the Ara OB1 association,
supplemented by ground-based near-IR images.
The main results of this study
are summarized as follows:

\begin{enumerate}

\item {\em Chandra}  detected 43 X-ray sources 
in a $\approx$2$'$ $\times$ 2$'$ region centered on
the O stars HD 150135 and HD 150136. The
strong X-ray emission previously seen by 
{\em ROSAT} is now attributed to separate but
unequal contributions from HD 150135 and
HD 150136. The X-ray luminosity of HD 150136
is exceptional (log L$_{\rm X}$ = 33.39 ergs s$^{-1}$) 
making it one of the most X-ray luminous O stars known. \\

\item All 43 X-ray sources in the central cluster region 
have near-IR counterparts, but only 
eleven have optical identifications. One of these,
the early B star CD$-$48 11069, shows an unusually
hard photon energy distribution,  raising the possibility
that its X-ray emission arises in a low-mass companion.
The population of fainter X-ray sources surrounding
the central O stars is characterized by moderately
hard emission with mean photon energies 
$\langle$E$\rangle$ $\approx$ 2 keV and about one-third
are variable. The variable objects are promising 
candidates for low-mass pre-main-sequence stars in
the cluster, but membership remains to be determined.
The available JHK photometry suggest that the fraction
of X-ray sources in the central cluster region with
near-IR excesses is low. Further searches for 
excesses in the mid-IR are needed. \\

\item High-quality X-ray grating spectra of 
 HD 150135 and HD 150136 show similar, but not
identical, spectral properties. In both stars,
the emission is dominated by cool plasma 
(kT  $\approx$ 0.3 keV). The emission lines
show no significant centroid shifts and
are moderately broadened to HWHM $\approx$
$v_{\infty}$/3. Broadened lines and cool plasma
implicate wind shock emission as the origin
of the X-rays. Forbidden lines in He-like 
triplets are strongly suppressed by the
HD 150136 UV radiation field, implying
that some high-temperature lines such as
S XV form in the wind-acceleration zone
within a few radii of the star. Optical
depth and line broadening considerations
hint that the wind may be clumped. \\

\item The exceptionally high X-ray luminosity
of HD 150136, along with  nonthermal radio
emission, suggest that colliding wind shocks
may be present in this closely-spaced O3 $+$ O6V
binary. However, simplified numerical simulations
show that a CW shock alone cannot account for
all of the observed X-ray properties. If CW
emission is present, then it is supplemented
by another cool X-ray component, which 
likely originates in the shocked winds of 
the individual O stars. \\

\item An unexpected result is that the X-ray
emission of HD 150136  is slowly variable on
a timescale of $\la$1 day. The variability
is due to a change in the emission measure,
with little or no change in X-ray temperature,
suggesting that it is an occultation effect.
The mechanism responsible for the occultation
is not yet known, but self-occultation due
to stellar rotation or partial occultation of
the X-ray emitting region by a putative 
third star in the HD 150136 system are 
possibilities. Further observational work,
including long-term X-ray monitoring, will
be needed to identify the origin of the
X-ray variability.

\end{enumerate}

\section*{Acknowledgments}

This research was supported by NASA/SAO grant GO2-3025X.
This publication makes use of data products from the 
Two Micron All Sky Survey (2MASS), which is a joint project of
the Univ. of Massachusetts and the Infrared Processing
and Analysis Center / California Institute of Technology,
funded by the National Aeronautics and Space Administration
and the National Science Foundation. The results presented
here are partially based on observations made at 
Observat\'orio do Pico dos Dias/LNA.  SZ and FP acknowledge
partial financial support from the Bulgarian Academy of
Sciences - Consiglio Nazionale delle Ricerche bilateral 
cooperation program. CB thanks FAPESP and Inst.
do Mil\^enio for financial suport.

\newpage

\clearpage

\begin{table*}
 \centering
 \begin{minipage}{140mm}
  \caption{Stellar Properties}
  \begin{tabular}{@{}llll@{}}
  \hline
   Property     & HD 150135     & HD 150136  & Reference$^a$ \\  
                &               &            &         \\
  \hline
Spectral type & O6.5V $+$ ?   & O3 $+$ O6V & (1),(3)   \\
Binary        & possible (SB1)& yes (SB2)  & (2),(4)   \\
P$_{orb}$ (d) & ...           & 2.662      & (3)        \\
V (mag)       & 6.89          & 5.62       & (1)        \\
E(B$-$V)      & 0.49          & 0.48       & (1)   \\
A$_{V}$ (mag) & 1.7           & 1.7        & (1)    \\
M$_{V}$ (mag) & $-$5.33       & $-$6.56    & (1)        \\
$v_{\infty}$ (km s$^{-1}$) & [2455]$^b$    & 3160   & (5)  \\
Distance (kpc)$^c$            & 1.32 $\pm$ 0.12   & 1.32 $\pm$ 0.12  & (1)  \\
F$_{X}$ (10$^{-12}$ ergs cm$^{-2}$ s$^{-1}$)$^d$ & 0.5 (3.0) & 3.3 (12.1) & (6) \\
log L$_{X}$ (ergs s$^{-1}$)$^e$  & 32.78      & 33.39 & (6)  \\
log (L$_{X}$/L$_{bol}$)$^f$      & $-$6.3        & $-$6.4& (6)       \\
\hline
\end{tabular}

$^a${\em Refs.}: 
(1) Herbst \& Havlen (1977)~(2) Arnal et al. (1988)~ \\
(3) Niemela \& Gamen (2005)~(4) Garmany, Conti, \& Massey (1980)~ \\
(5) Prinja et al. (1990)~(6) this work \\
$^b$ Typical value for O6.5V (Prinja et al. 1990) \\
$^c$ To association (HH77) \\
$^d$
X-ray flux (F$_{X}$) is the absorbed value in the 0.5 - 6 keV
band followed in parentheses by the unabsorbed value. Fluxes
are from best-fit models of {\em Chandra} spectra using an absorption column
density N$_{\rm H}$ = 4 $\times$ 10$^{21}$ cm$^{-2}$. \\
$^e$X-ray luminosity (L$_{X}$) is the unabsorbed value in the
0.5 - 6 keV band at a distance of 1.3 kpc. \\ 
$^f$L$_{bol}$ is from Vacca et al. (1996). The undetermined
luminosity class of the primary in the HD 150136 system
(O3 $+$ O6V) leads to an uncertainty of $\pm$0.1 dex
in  L$_{bol}$. \\
\end{minipage}
\end{table*}

\clearpage


\begin{table*}
 \centering
 \begin{minipage}{140mm}
  \caption{X-Ray Sources in the NGC 6193 Core Region}
  \begin{tabular}{@{}lllllll@{}}
  \hline
No. & Name              & RA           & Dec.          & Net counts & $<$E$>$& Identification (offset) \\
    &                   & (J2000.0)    & (J2000.0)     &            & (keV)  & ~~~~~~~~~~~~~~~~~~(arcsec)     \\
    &                   &              &               &            &        &           \\
  \hline
1   & J1641139$-$484553 & 16 41 13.99 & $-$48 45 53.30 &100 $\pm$ 10      & 2.13 & 2M 16411400-4845528 (0.5) \\
2   & J1641154$-$484609 & 16 41 15.43 & $-$48 46 09.72 & 10 $\pm$ 3       & 2.60 & 2M 16411539-4846092 (0.6) \\
3   & J1641167$-$484516 & 16 41 16.72 & $-$48 45 16.74 & 26 $\pm$ 5       & 1.92 & 2M 16411672-4845165 (0.2) \\
4   & J1641169$-$484546 & 16 41 16.95 & $-$48 45 46.08 & 44 $\pm$ 7 (v)   & 1.50 & PD 16411695-4845456 (0.4) \\
5   & J1641170$-$484603 & 16 41 17.07 & $-$48 46 03.76 &118 $\pm$ 11      & 1.84 & 2M 16411706-4846034 (0.3) \\
6   & J1641175$-$484601 & 16 41 17.54 & $-$48 46 01.11 & 24 $\pm$ 5       & 2.24 & 2M 16411756-4846008 (0.3) \\
7   & J1641175$-$484507 & 16 41 17.54 & $-$48 45 07.36 & 16 $\pm$ 4       & 1.85 & 2M 16411752-4845071 (0.3); GSC (ns) \\
8   & J1641177$-$484519 & 16 41 17.78 & $-$48 45 19.20 &  4 $\pm$ 2       & 2.89 & 2M 16411779-4845190 (0.2) \\
9   & J1641181$-$484535 & 16 41 18.14 & $-$48 45 35.04 & 13 $\pm$ 4 (v)   & 1.80 & PD 16411814-4845345 (0.5) \\
10  & J1641184$-$484628 & 16 41 18.41 & $-$48 46 28.36 &  6 $\pm$ 3       & 1.58 & 2M 16411843-4846281 (0.4) \\
11  & J1641184$-$484536 & 16 41 18.46 & $-$48 45 36.78 & 15 $\pm$ 4       & 1.97 & PD 16411846-4845362 (0.6) \\
12  & J1641185$-$484500 & 16 41 18.53 & $-$48 45 00.73 & 41 $\pm$ 6 (v)   & 2.25 & 2M 16411852-4845004 (0.4) \\
13  & J1641190$-$484602 & 16 41 19.06 & $-$48 46 02.52 & 58 $\pm$ 8       & 1.94 & 2M 16411905-4846024 (0.2) \\
14  & J1641194$-$484519 & 16 41 19.46 & $-$48 45 19.58 & 145 $\pm$ 12 (v) & 2.38 & 2M 16411944-4845194 (0.3) \\
15  & J1641194$-$484547 & 16 41 19.46 & $-$48 45 47.78 & 888 $\pm$ 30     & 1.22 & 2M 16411945-4845475 (0.3); HD 150135 \\
16  & J1641194$-$484535 & 16 41 19.47 & $-$48 45 35.15 &   4 $\pm$ 2      & 1.56 & PD 16411945-4845349 (0.2) \\
17  & J1641200$-$484554 & 16 41 20.08 & $-$48 45 54.00 &  21 $\pm$ 5      & 2.00 & PD 16412006-4845535 (0.5) \\
18  & J1641201$-$484523 & 16 41 20.10 & $-$48 45 23.22 &  19 $\pm$ 5      & 2.03 & 2M 16412001-4845231 (0.9) \\
19  & J1641201$-$484515 & 16 41 20.15 & $-$48 45 15.27 &  27 $\pm$ 5      & 1.89 & 2M 16412012-4845150 (0.4); GSC (ns)  \\
20  & J1641204$-$484546 & 16 41 20.44 & $-$48 45 46.93 &7293 $\pm$ 86 (v) & 1.42 & 2M 16412042-4845466 (0.4); HD 150136 \\
21  & J1641205$-$484536 & 16 41 20.52 & $-$48 45 36.50 &   6 $\pm$ 3      & 2.46 & PD 16412051-4845359 (0.6) \\
22  & J1641209$-$484559 & 16 41 20.92 & $-$48 45 59.43 &   8 $\pm$ 3      & 1.71 & 2M 16412089-4845590 (0.5) \\
23  & J1641210$-$484642 & 16 41 21.02 & $-$48 46 42.00 &  61 $\pm$ 8 (v)  & 2.32 & 2M 16412099-4846418 (0.3) \\
24  & J1641210$-$484521 & 16 41 21.07 & $-$48 45 21.86 & 45  $\pm$ 7 (v)  & 1.63 & 2M 16412106-4845217 (0.2) \\
25  & J1641217$-$484641 & 16 41 21.75 & $-$48 46 41.26 & 137 $\pm$ 12 (v) & 2.43 & 2M 16412172-4846411 (0.4) \\
26a & J1641221$-$484457 & 16 41 22.11 & $-$48 44 57.22 & 224 $\pm$ 15 (v) & 2.09 & 2M 16412210-4844570 (0.3); CD$-$48 11069 \\
26b & J1641222$-$484454 & 16 41 22.25 & $-$48 44 54.80 &  17 $\pm$ 4      & 3.91 & PD 16412227-4844547 (0.3) \\ 
27  & J1641222$-$484604 & 16 41 22.26 & $-$48 46 04.32 &  12 $\pm$ 4      & 1.99 & 2M 16412220-4846044 (0.6) \\
28  & J1641223$-$484628 & 16 41 22.36 & $-$48 46 28.28 &  17 $\pm$ 4      & 1.81 & 2M 16412236-4846278 (0.5) \\
29  & J1641224$-$484537 & 16 41 22.42 & $-$48 45 37.08 &  72 $\pm$ 9      & 1.94 & 2M 16412240-4845369 (0.3) \\
30  & J1641225$-$484524 & 16 41 22.51 & $-$48 45 24.42 &  22 $\pm$ 5 (v)  & 2.17 & 2M 16412247-4845241 (0.5); GSC (ns) \\
31  & J1641225$-$484559 & 16 41 22.59 & $-$48 45 59.68 &  15 $\pm$ 4      & 2.39 & 2M 16412258-4845594 (0.3) \\
32  & J1641227$-$484535 & 16 41 22.77 & $-$48 45 35.32 &  20 $\pm$ 5      & 2.04 & 2M 16412276-4845350 (0.3) \\
33  & J1641232$-$484449 & 16 41 23.27 & $-$48 44 49.70 &  16 $\pm$ 4      & 1.78 & 2M 16412327-4844494 (0.3); GSC (s) R = 16.5 \\
34  & J1641234$-$484604 & 16 41 23.46 & $-$48 46 04.37 & 144 $\pm$ 12 (v) & 2.21 & 2M 16412343-4846041 (0.4) \\
35  & J1641238$-$484520 & 16 41 23.84 & $-$48 45 20.85 &  46 $\pm$ 7  (v) & 1.82 & 2M 16412383-4845206 (0.2); GSC (ns)  \\
36  & J1641242$-$484543 & 16 41 24.26 & $-$48 45 43.83 &  32 $\pm$ 6      & 1.87 & 2M 16412425-4845436 (0.2) \\
37  & J1641244$-$484552 & 16 41 24.40 & $-$48 45 52.78 &  52 $\pm$ 7      & 1.99 & 2M 16412440-4845524 (0.3); NGC6193-9 \\
38  & J1641247$-$484534 & 16 41 24.73 & $-$48 45 34.68 &  25 $\pm$ 5      & 1.67 & 2M 16412471-4845345 (0.3) \\
39  & J1641247$-$484545 & 16 41 24.75 & $-$48 45 45.13 &  15 $\pm$ 4      & 1.51 & 2M 16412475-4845452 (0.1) \\
40  & J1641256$-$484619 & 16 41 25.60 & $-$48 46 19.28 &  32 $\pm$ 6  (v) & 1.60 & 2M 16412558-4846189 (0.4); GSC (s) R = 16.7\\
41  & J1641258$-$484514 & 16 41 25.88 & $-$48 45 14.51 &  44 $\pm$ 7      & 0.87 & 2M 16412586-4845142 (0.3); CD$-$48 11071 \\
42  & J1641262$-$484548 & 16 41 26.28 & $-$48 45 48.91 &  19 $\pm$ 4  (v) & 1.76 & 2M 16412625-4845487 (0.4) \\
\hline
\end{tabular}
{\em Notes}: Sources detected by the CIAO {\em wavdetect} algorithm within a 2.1$'$ $ \times$ 2.1$'$ 
region centered on HD 150136 are included in this table. Net counts from {\em wavdetect}
were accumulated in ACIS-S 0th order images during an exposure live time of 90,337 secs 
and are background subtracted. Dispersed counts in the MEG and HEG grating arms are not included
in the net counts quoted above for HD 150135 and HD 150136. Sources 8 and 16 are marginal X-ray detections.
Sources 26a,b are a close pair separated by 2.8$''$.
A ``(v)'' following net counts indicates that the source is likely variable with a probability of
constant count rate P$_{\rm const}$ $<$ 0.05 as determined from the KS statistic using source
events in the 0.5 - 7.0 keV range.
$<$E$>$ is the mean photon energy using source events in the 0.5 - 7.0 keV range.
Infrared identifications are from 2MASS (2M) and Pico dos Dias (PD) images. The offset between X-ray
and IR positions is given in parentheses. Sources with optical counterparts within 1$''$ of the
X-ray position in the  {\em HST} Guide Star Catalog are denoted as GSC (s) if the GSC classification
is stellar and GSC (ns) if non-stellar. The GSC R magnitude is given if stellar.
\end{minipage}
\end{table*}

\clearpage

\begin{table*}
 \centering
 \begin{minipage}{140mm}
  \caption{Chandra Spectral Fit Results}
  \begin{tabular}{@{}lll@{}}
  \hline
Parameter                             & HD 150135              & HD 150136    \\
                                      &                        &              \\
\hline
Spectrum                              & MEG1                   & MEG1         \\
Net counts (MEG)                      & 1098                   & 8780         \\
Model                                 & 2T bvapec              & 2T bvapec     \\
N$_{\rm H}$ (10$^{21}$ cm$^{-2}$)     & [4.0]                  & [4.0]         \\
kT$_{1}$ (keV)                        & 0.23 $\pm$ 0.10        & 0.35 $\pm$ 0.04       \\ 
kT$_{2}$ (keV)                        & 0.57 $\pm$ 0.25        & 1.00 $\pm$ 0.07      \\
EM$_{1}$ (10$^{56}$ cm$^{-3}$)        & 0.90                   & 2.29       \\
EM$_{2}$ (10$^{56}$ cm$^{-3}$)        & 0.13                   & 0.62    \\
FWHM (km s$^{-1}$)         & 2423 $\pm$ 1692        & 2547 $\pm$ 182    \\
Abundances                            & varied$^a$             & varied$^b$             \\
$\chi^2$/dof                          & 9.5/44                 & 242.1/383                \\
$\chi^2_{red}$                        & 0.22                   & 0.63                  \\
\hline
\end{tabular}

$^a$Best-fit HD 150135 abundances relative to the solar values of
Anders \& Grevesse (1989) and 90\% confidence errors were:
O = 0.45 ($+$0.75, $-$0.39), Ne = 0.40 ($\pm$ 0.31), Mg = 0.33 ($\pm$ 0.30),
Si = 0.47 ($+$0.63, $-$0.37), Fe = 0.35 ($\pm$ 0.25); all other elements 
fixed at solar values. \\
$^b$Best-fit HD 150136 abundances relative to the solar values
of Anders \& Grevesse (1989) and 90\% confidence errors were:
O = 0.35 ($\pm$ 0.20), Ne = 0.33 ($\pm$ 0.09), Mg = 0.56 ($\pm$ 0.11),
Si = 0.77 ($\pm$ 0.12), S = 1.10 ($\pm$ 0.44), Fe = 0.20 ($\pm$ 0.06); all other elements
fixed at solar values. \\
{\em Notes}: 
Based on  fits of the MEG 1st order  spectra binned to a minimum of 
20 counts per bin (Figs. 8 - 9) using an absorbed  2T {\em bvapec} optically thin plasma 
model in XSPEC v. 11.3.1. The tabulated parameters
are absorption column density (N$_{\rm H}$), plasma temperature (kT),
emission measure (EM), and the best-fit line width (FWHM).
Errors are 90\% confidence.
N$_{\rm H}$ = 4.0 $\times$ 10$^{21}$ cm$^{-2}$ was held fixed during fitting 
at the value derived from E(B-V) = 0.48 - 0.49 and R = A$_{V}$/E(B-V) = 3.5
(HH77) using the conversion formula of Gorenstein (1975). If N$_{\rm H}$ is varied,
the higher S/N spectrum of HD 150136 gives a 90\% confidence range
N$_{\rm H}$ = [3.4 - 6.0] $\times$ 10$^{21}$ cm$^{-2}$. 

\end{minipage}
\end{table*}

\clearpage
\newpage

\begin{table*}
 \centering
 \begin{minipage}{140mm}
  \caption{HD 150136 X-ray Line Properties}
  \begin{tabular}{@{}llllllllllll@{}}
  \hline
Ion    & Inst. & $\lambda_{lab}$ & $\lambda_{obs}$ & $\Delta$$\lambda$ & FWHM     &FWHM   & Flux  & R & G &  T$_{max}$ & R$_{c}$  \\
       &         & (\AA)           & (\AA)           & (m\AA)            & (m\AA)   & (km/s)&       &   &  & (K)       & (R$_{*}$) \\
       &         &                 &                 &                   &          &       &       &   &  &           &      \\
  \hline
S XV   & HEG& 5.0387   & 5.0351 $\pm$ 0.011  & $-$3.6  & 33$^{+28}_{-...}$    & 1965$^{+833}_{-...}$  & 1.26$^{+1.27}_{-0.94}$ 
       & 1.6$^{+0.5}_{-1.5}$  & 0.9$^{+3.8}_{-0.8}$ & 7.2  & 1.7  \\
Si XIV & HEG& 6.1804   & 6.1816 $\pm$ 0.005  & $+$1.2  & 38$^{+11}_{-8}$      & 1844$^{+534}_{-388}$  & 1.02$^{+0.22}_{-0.22}$
       & ...                  & ...                 & 7.2  & ... \\
Si XIII& HEG  & 6.6479   & 6.6482 $\pm$ 0.006  & $+$0.3  & 43$^{+12}_{-8}$    & 1940$^{+541}_{-361}$  & 4.01$^{+0.85}_{-1.70}$
       & 1.0$^{+0.9}_{-0.4}$                   & 0.9$^{+0.4}_{-0.3}$          & 7.0 & 3.9    \\
Mg XII & HEG  & 8.4192   & 8.4163 $\pm$ 0.004  & $-$2.9  & 42$^{+9}_{-7}$     & 1496$^{+321}_{-249}$  & 1.44$^{+0.29}_{-0.29}$
       & ...                  & ...                 & 7.0   & ... \\
Mg XI  & HEG  & 9.1687   & 9.1688 $\pm$ 0.005  & $+$0.1  & 76$^{+16}_{-11}$   & 2487$^{+524}_{-360}$  & 7.15$^{+1.35}_{-1.35}$
       & 0.6$^{+0.4}_{-0.3}$                   & 0.6$^{+0.3}_{-0.5}$          & 6.8 & 10.1    \\
Ne X   & HEG  & 12.1321  &12.1288 $\pm$ 0.010  & $-$3.3  & 92$^{+18}_{-15}$   & 2275$^{+445}_{-371}$  & 6.80$^{+1.17}_{-1.17}$
       & ...                                   & ...  & 6.8   & ... \\
Ne IX  & MEG  & 13.4473  &             ......  & ......  &105$^{+15}_{-12}$   & 2342$^{+335}_{-268}$  & 10.3$^{+1.70}_{-1.70}$
       & 0.2$^{+0.2}_{-0.1}$                   & 1.0$^{+0.3}_{-0.3}$          & 6.6 & 30.7    \\
O VIII & MEG  & 18.9671  &18.9692 $\pm$ 0.012  & $+$2.1  &176$^{+21}_{-21}$   & 2784$^{+332}_{-332}$  & 11.2$^{+1.50}_{-1.50}$
       & ...                                   & ...                          & 6.5   & ... \\
O VII  & MEG  & 21.6015  &21.5998 $\pm$ 0.032  & $-$1.7  &121$^{+53}_{-32}$   & 1680$^{+736}_{-444}$  & 6.70$^{+4.14}_{-3.16}$
       & $\leq$0.1$^{+0.4}_{-0.1}$             & 1.2$^{+2.9}_{-0.8}$          & 6.3  & 116. \\
\hline 
\end{tabular}

{\em Notes}: Data are from Gaussian fits of first-order grating spectra.
$\Delta$$\lambda$ =  $\lambda_{obs}$ $-$ $\lambda_{lab}$, where 
$\lambda_{lab}$ is the rest wavelength from ATOMDB v. 1.3.1 and $\lambda_{obs}$ is the observed
(best-fit) wavelength.   Line flux units are
10$^{-5}$ ph cm$^{-2}$ s$^{-1}$. Errors are $\pm$1$\sigma$. For He-like triplets, 
$\lambda_{obs}$ is for the resonance ($r$) line, and fluxes are the 
sum of contributions from  the 
resonance ($r$), intercombination ($i$), and forbidden ($f$) lines.
Flux ratios are R = f/i and G = (f $+$ i)/r. In the low-density limit,
R $\rightarrow$ R$_{0}$, where  R$_{0}$ = 2.04 (S XV), 2.51 (Si XIII),
3.03 (Mg XI), 3.17 (Ne IX), 3.85 (O VII) (Blumenthal et al. 1972).
For Ne IX, the wavelength was held fixed
at  $\lambda_{lab}$ during fitting do to blending from nearby Fe lines.
The O VII line is faint and measurements have large uncertainties.
T$_{max}$ is the log temperature of maximum line power from ATOMDB.
R$_{c}$ for He-like triplets is the critical radius (in units of stellar radii) 
where the photoexcitation rate from the upper level of the $f$ transition is
equal to the spontaneous decay from that level, computed using the 
procedure of Kahn et al. (2001).  Quoted values of R$_{c}$
assume a spectral type of O3V  and blackbody spectra with T$_{eff}$ = 51230 K 
(Vacca et al. 1996). The R$_{c}$ values for O3I are $\approx$1\% less.

\end{minipage}
\end{table*}

\clearpage

\newpage

\begin{table*}
 \centering
 \begin{minipage}{140mm}
  \caption{Adopted O Star Parameters}
  \begin{tabular}{@{}lccll@{}}
  \hline
Spectral type            & M$^a$         & R$^a$        & log $\dot{\rm M}$      & v$_{\infty}$$^c$    \\
                         & (M$_{\odot}$) & (R$_{\odot}$)& (M$_{\odot}$ yr$^{-1}$)& (km s$^{-1}$)   \\
                         &               &              &                        &              \\
\hline
O3I                      & 64.4          & 17.8         & $-$4.9$^d$             & 3150         \\
O3V                      & 51.3          & 13.2         & $-$5.7$^b$             & 3190         \\
O6V                      & 33.1          & 10.7         & $-$6.0$^d$             & 2570         \\
O6.5V                    & 30.8          & 10.3         & $-$6.2$^b$             & 2455         \\
\hline
\end{tabular}

$^a$Vacca et al. 1996 \\
$^b$Garmany et al. 1981  \\
$^c$Prinja et al. 1990   \\
$^d$Lamers \& Leitherer 1993 \\

\end{minipage}
\end{table*}




\clearpage
\newpage

\begin{figure*}
\begin{center}
\includegraphics[width=114mm,angle=270.]{fig1a.ps}
\vspace*{0.5cm}
\includegraphics[width=105mm,angle=0.]{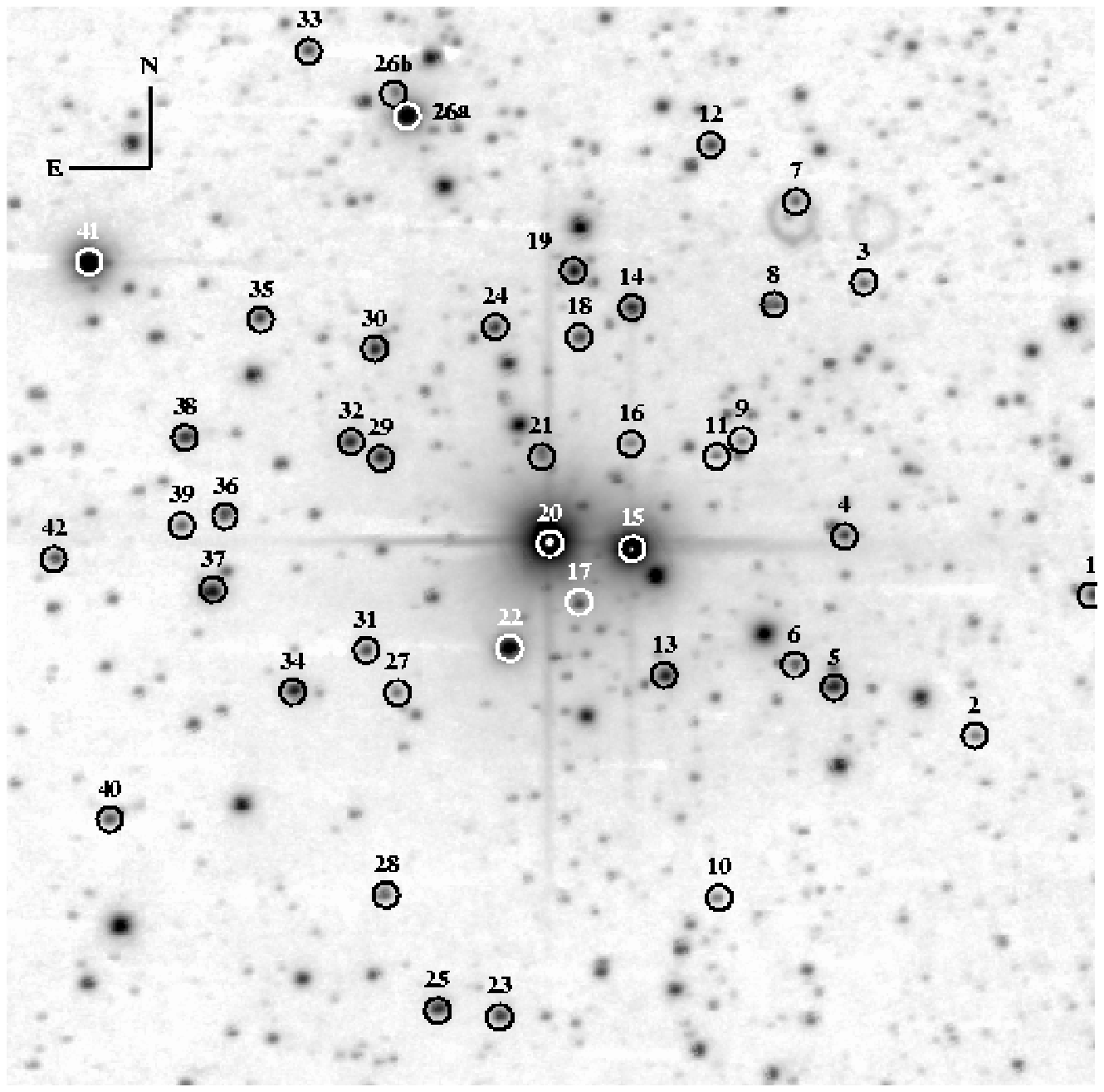}
\end{center}
\caption{{\em Top}: Full resolution Chandra ACIS-S 0th order image of the 
         central 2.1$'$ $\times$ 2.1$'$ 
         region of NGC 6193 in the 0.5 - 7.0 keV energy range. The pixel size
         is 0.492$''$ and coordinates are J2000.
         Boxes enclose seven X-ray sources which have stellar
         counterparts within 1$''$ of the X-ray position (Table 2).
         {\em Bottom}: H-band image of the central region of NGC 6193 obtained with the
                       Pico dos Dias 1.6 m telescope. The numbered sources are X-ray
                       detections (Table 2). The faint rings near source 7 are
                       filter reflection artifacts.}
\end{figure*}

\clearpage
\newpage

\begin{figure*}
\includegraphics[width=80mm,angle=270.]{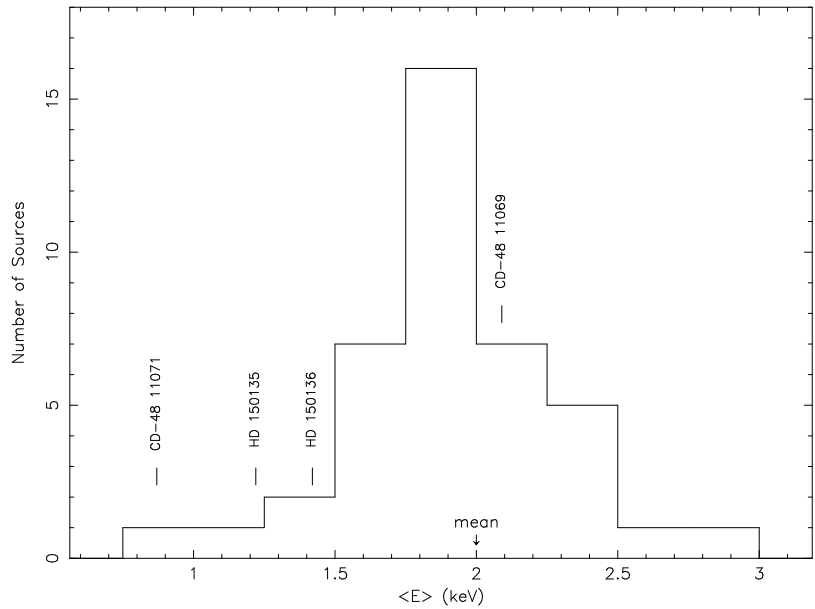}
\caption{Distribution of the mean photon energy $\langle$E$\rangle$
         for the 43 X-ray sources detected in the central region of
         NGC 6193. Three OB stars shown have soft energy distributions,
         but the B0-1 star CD $-$48 11069 has an above average 
         $\langle$E$\rangle$ that may be due to a low-mass companion. }
\end{figure*}

\clearpage
\newpage

\begin{figure*}
\includegraphics[width=80mm,angle=270.]{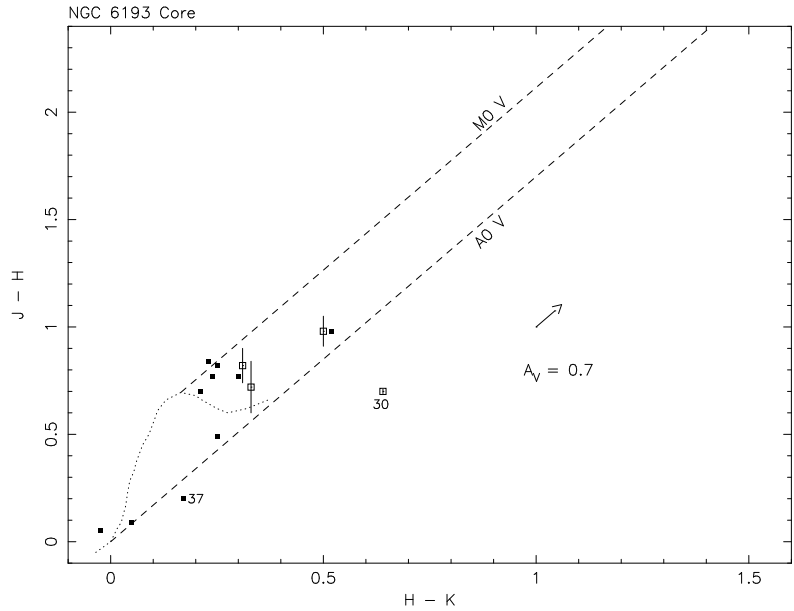}
\caption{Near-IR color-color diagram for central region of
NGC 6193. Solid squares show 2MASS colors of 10 X-ray sources
whose 2MASS JHK magnitudes do not have confusion flags
(X-ray sources 1, 5, 23, 25, 26a, 34, 35, 37, 38, 41).
Open squares show 2MASS colors of four sources which 
have a confusion flag at J band, but not at H or K
(X-ray sources 2,3, 28, 30). The unreddened zero-age
main sequence is shown as a dotted line at lower left.
The two sloping dashed lines mark the approximate
reddening band for M0 V and A0 V stars using data
from Bessell \& Brett (1988) and Rieke \& Lebofsky (1985).
The reddening vector corresponds to  foreground reddening
E(B$-$V) = 0.2 mag and  
A$_{\rm v}$ = 3.5 $\times$ E(B$-$V) = 0.7 mag (HH77).}
\end{figure*}

\clearpage
\newpage

\begin{figure*}
\includegraphics[width=80mm,angle=270.]{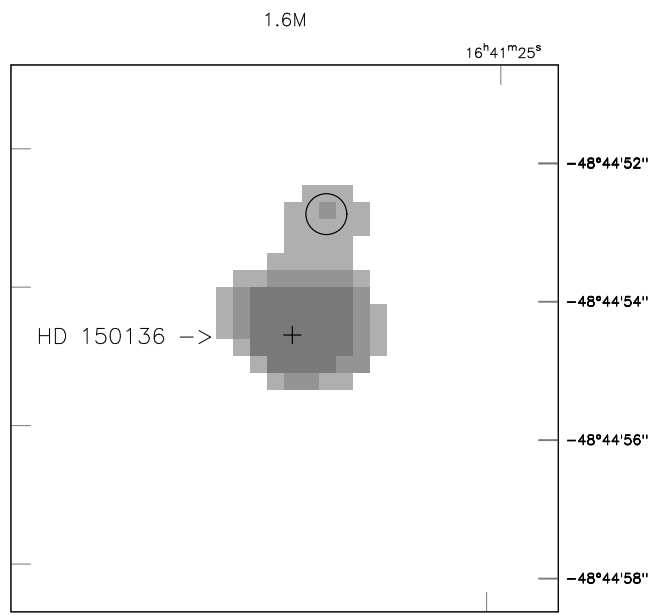}
\caption{Pico dos Dias 2.14 $\mu$m image of HD 150136,
showing the faint IR companion 1.7$''$ to its north.}
\end{figure*}

\clearpage
\newpage

\begin{figure*}
\includegraphics[width=80mm,angle=270.]{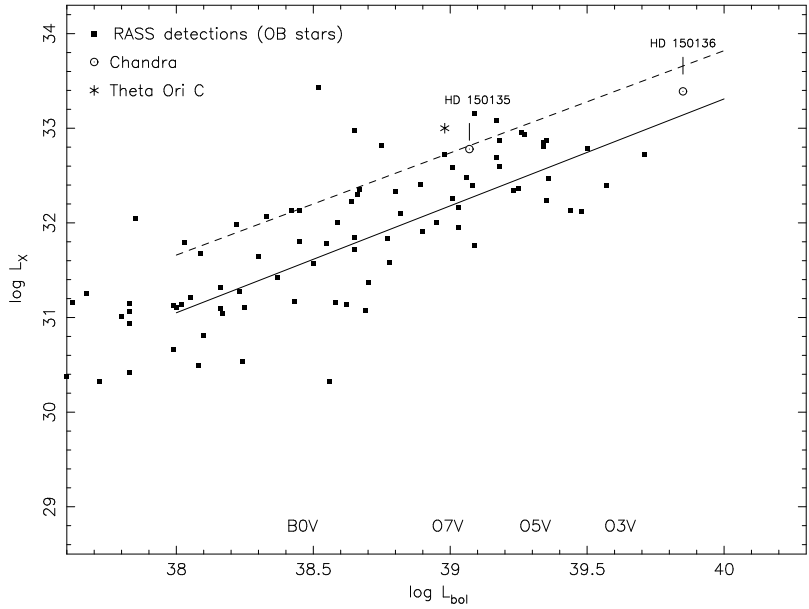}
\caption{L$_{X}$ versus L$_{bol}$ for luminous
OB stars detected in the  RASS (solid squares) as
catalogued by  Bergh\"{o}fer et al. (1996). Upper 
limits for undetected OB stars in the RASS are not 
shown. Circled symbols are {\em Chandra} data.
For comparison, the asterisk shows the young magnetic 
star $\theta^1$ Ori C (O7V) using  L$_{bol}$ from
Hillenbrand (1997) and L$_{X}$ from the {\em Chandra}
HETG analysis of Gagne et al. (2005).  
Solid line is the regression fit derived
for RASS OB stars, accounting for non-detections,
by Bergh\"{o}fer et al. (1997). Dashed line is
the regression fit for O stars determined in
the {\em Einstein} study of Sciortino et al. (1990).}
\end{figure*}

\clearpage
\newpage

\begin{figure*}
\includegraphics[width=80mm,angle=270.]{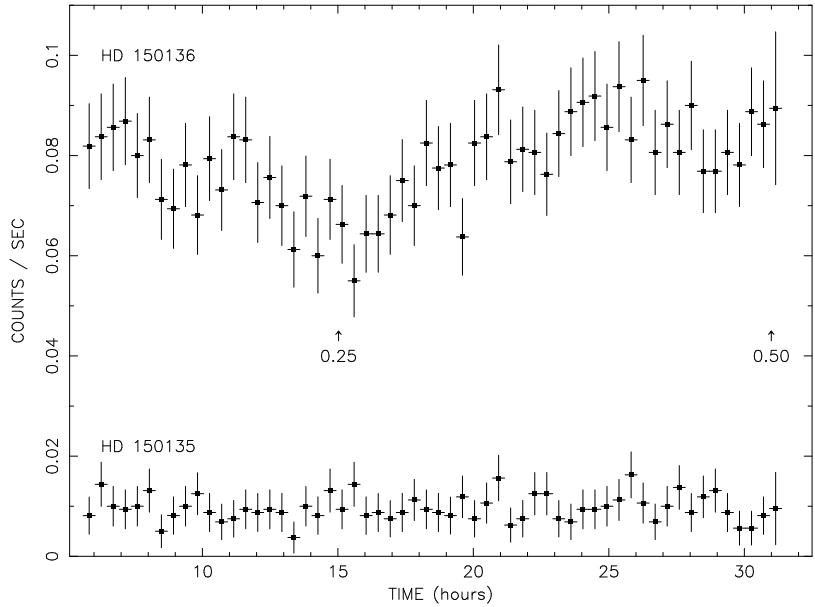}
\caption{Zeroth order {\em Chandra} ACIS-S light
curves of HD 150135 and HD 150136 in the 0.5 - 5 keV
band, binned at 1600 s intervals. Time (hours) is  referenced
to 00:00 UT on 27 June 2002. The observation began at
05:20 UT. Arrows mark HD 150136 orbital phases $\phi$ = 0.25 
and $\phi$ = 0.50  as determined from
the ephemeris of Niemela \& Gamen (2005). Absorption features
from a possible third star in the system appear in optical
spectra at $\phi$ = 0.25. The primary moves
in front at $\phi$ = 0.50.   
}
\end{figure*}

\clearpage
\newpage

\begin{figure*}
\includegraphics[width=80mm,angle=270.]{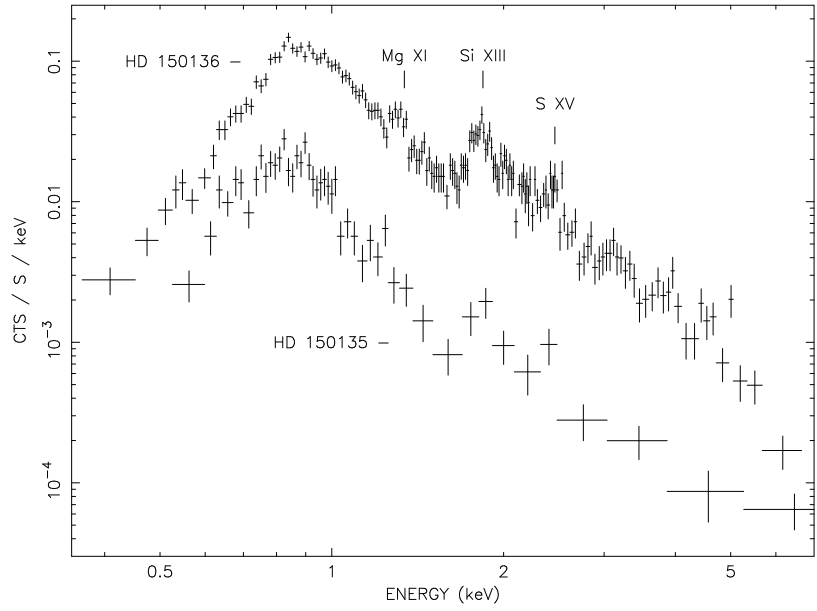}
\caption{Background-subtracted 0th order {\em Chandra} ACIS-S spectra
of 150135 and HD 150136.}
\end{figure*}

\clearpage
\newpage

\begin{figure*}
\includegraphics[width=80mm,angle=270.]{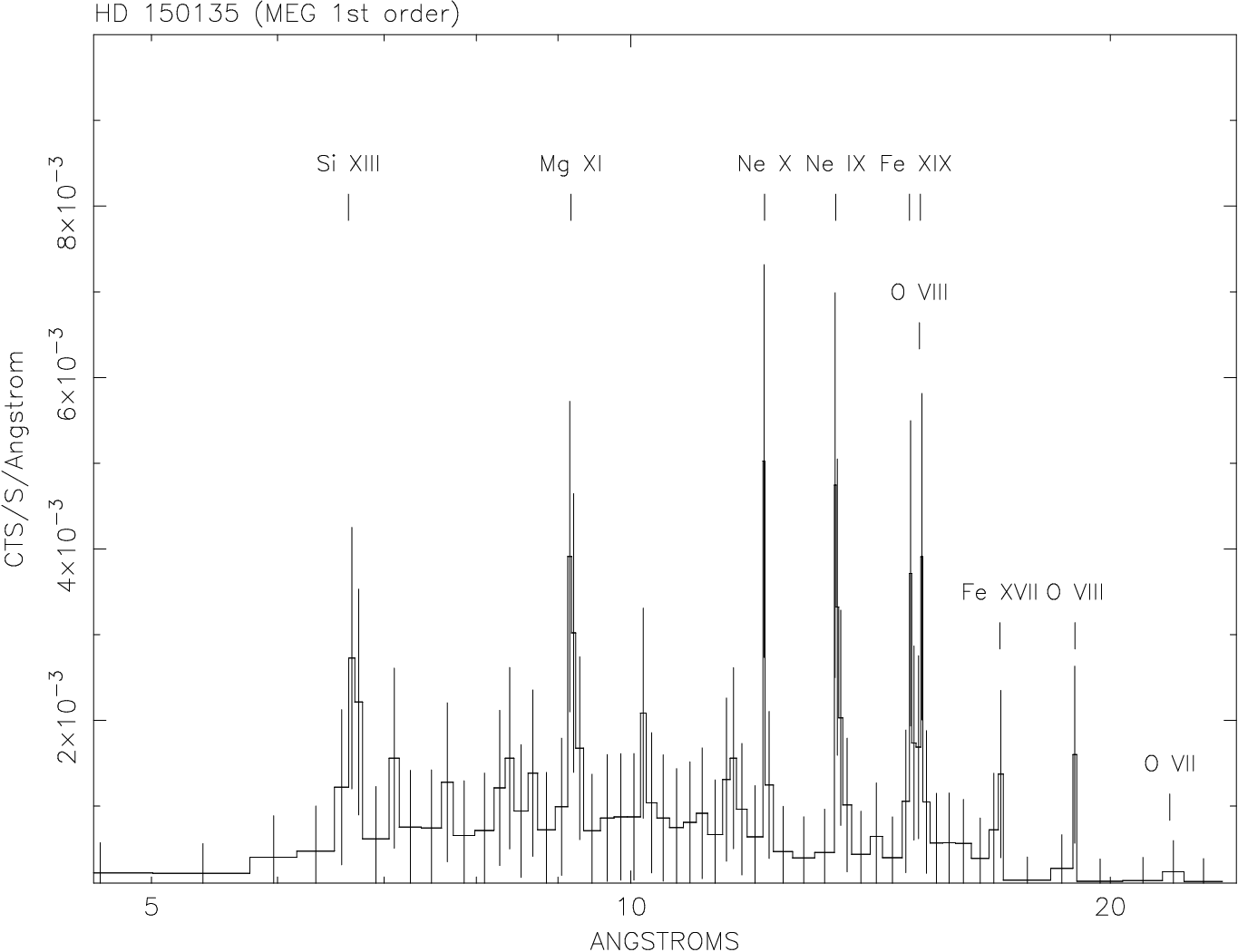}
\caption{ {\em Chandra} 1st order background-subtracted 
MEG spectrum of HD 150135. The $+$1 and $-1$ orders have
been co-added.}
\end{figure*}

\clearpage
\newpage

\begin{figure*}
\includegraphics[width=80mm,angle=270.]{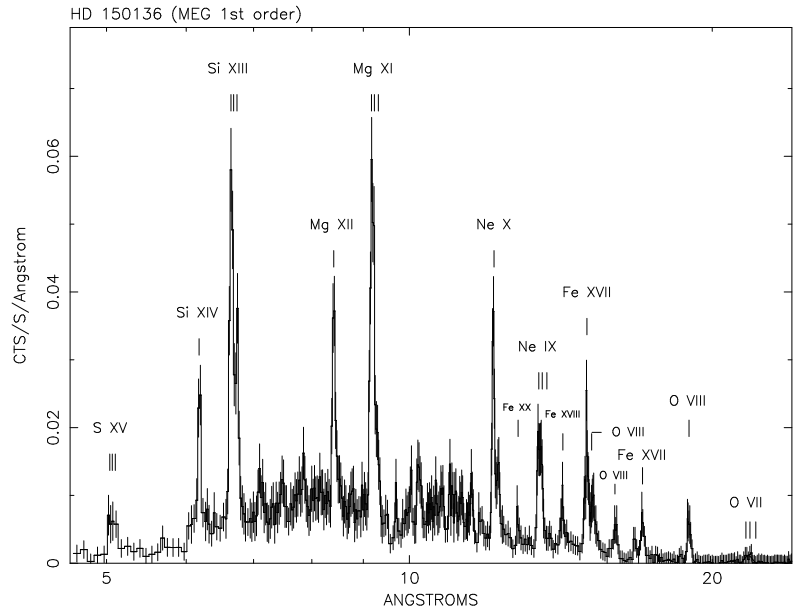}
\caption{{\em Chandra} 1st order background-subtracted
MEG spectrum of HD 150136. The $+$1 and $-1$ orders have
been co-added.}
\end{figure*}

\clearpage
\newpage

\begin{figure*}
\includegraphics[width=80mm,angle=270.]{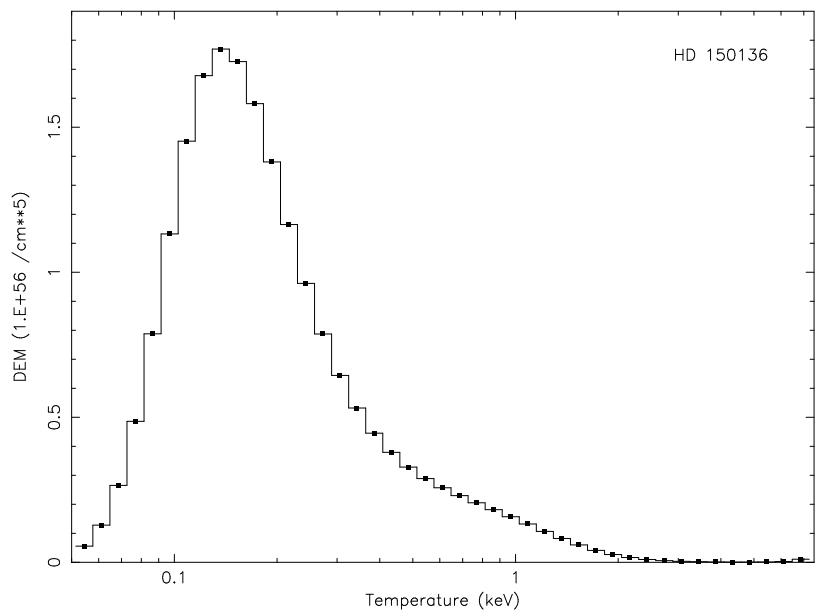}
\caption{Differential emission measure (DEM) distribution of HD 150136.
The DEM was derived using a Chebyshev polynomial algorithm (see text).}
\end{figure*}

\clearpage
\newpage

\begin{figure*}
\includegraphics[width=80mm,angle=270.]{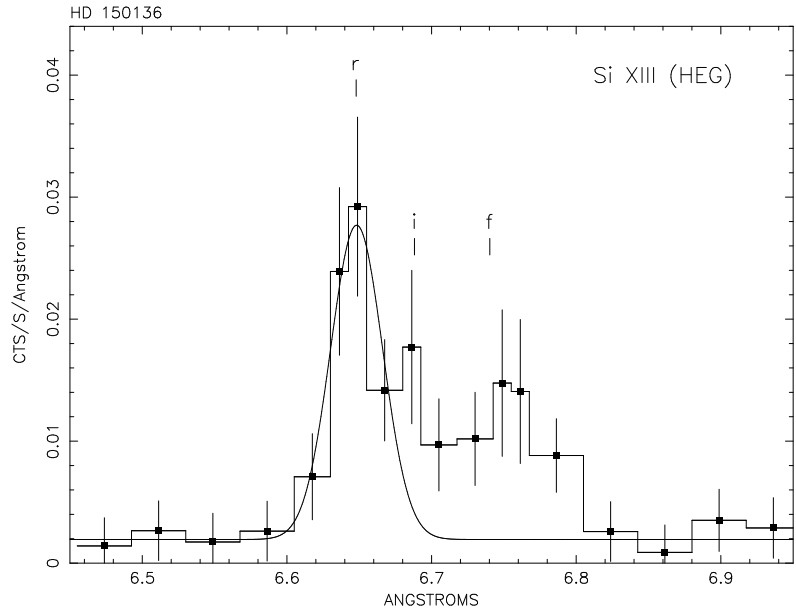}
\caption{Si XIII He-like triplet of HD 150136 from 
the background-subtracted 1st order {\em Chandra} HEG
spectrum. The $+1$ and $-1$ orders have been co-added. 
The Gaussian fit of the resonance ($r$)  line has
FWHM = 42.3 m\AA.}
\end{figure*}

\clearpage
\newpage

\begin{figure*}
\includegraphics[width=80mm,angle=270.]{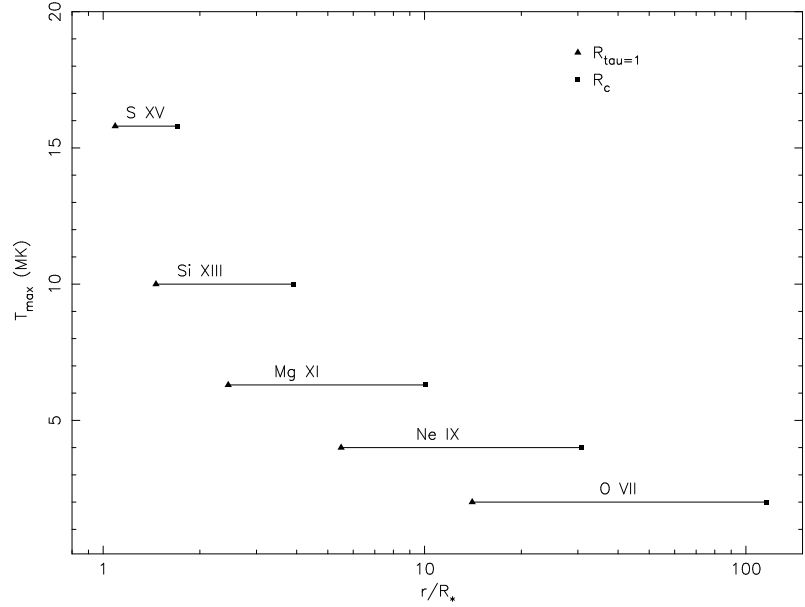}
\caption{Radius of optical depth unity (R$_{\rm tau = 1}$)
and critical photexcitation radius R$_{\rm c}$  (Table 4) for 
He-like triplets in HD 150136.  T$_{max}$ is the temperature at 
which line power is maximum.  The adopted stellar parameters are
for O3V  (Table 5). The value of R$_{\rm tau = 1}$ is 
calculated using solar abundance cross sections
(Balucinska-Church \& McCammon 1992) and assumes
a spherically symmetric homogeneous wind. 
The values of R$_{\rm tau = 1}$ for non-solar
abundances derived from X-ray spectral fits (Table 3)
are smaller by factors ranging from 6\% for S XV 
to 29\% for Ne IX and O VII.   The solid lines
connecting the two radii represent the  interval
over which the lines could form under the constraints
$r$ $>$ R$_{\rm tau = 1}$ and $r$ $<$ R$_{\rm c}$, where
the latter follows from UV suppressed R ratios (see text).}
\end{figure*}

\clearpage
\newpage

\bsp

\label{lastpage}

\end{document}